\documentclass[aps,pre,reprint,amsmath,amssymb,nofootinbib,floatfix]{revtex4-2}

\usepackage{graphicx}
\usepackage{bm}
\usepackage[colorlinks=true,citecolor=blue,linkcolor=blue,urlcolor=blue]{hyperref}
\makeatletter\let\auto@bib@innerbib\@empty\makeatother

\begin{document}

\title{Conviction versus Conformity in a Random-Field Ising Electorate:\\ Antagonistic Ties, Network Density, and a Parameter-Free Prediction}

\author{Natalia Romero}
\affiliation{Machine Perception and Cognitive Robotics Laboratory, Florida Atlantic University\\Natural and Artificial Intelligence Institute}

\begin{abstract}
We study the zero-temperature random-field Ising model with $\pm J$ bond disorder on sparse graphs as a
model of an electorate: survey-derived, trimodal fields compete with social couplings, a fraction $\alpha$
of which are antagonistic. With exact ground states up to $N=180$ (over $78\,000$ optimizations, certified
except in the densest frustrated cases) and validated heuristics to $N=720$, we locate the crossover
$r_h^*$ at which the field carries half of the ground-state energy, close to where half of the voters held
against their own field have switched ($0.93\,r_h^*$ for $\alpha\le0.2$). Our main test is a zero-temperature cavity theory
solved on the graph ensemble with no fitted parameter: at seven densities not otherwise used, predicted before
their exact ground states were computed, it placed $r_h^*$ within $10\%$ in 19 of 21 cells (median error
$3\%$). Its errors are unsigned at $\alpha=0.05$ and $0.5$ but all negative at $\alpha=0.2$, where it predicts
the coupling energy correctly but a crossover that is early and, at high density, too sharp, and where its
replica-symmetric solution is unstable in six of seven cells, consistent with replica-symmetry breaking at
intermediate frustration. Empirically $r_h^*\approx0.91\,e_J(0)$, with $e_J(0)$ the coupling energy of the
zero-field ground state, inside a rigorous per-instance factor-of-two window, so the known $\pm J$ energy laws
carry over: $r_h^*\propto\langle k\rangle^{\gamma}$ with $\gamma=1.01$--$1.09$ for $\alpha=0.05$ and
$0.50$--$0.57$ for $\alpha\ge0.5$ ($\langle k\rangle=3$--$28$; five graph families agree at matched density).
At intermediate frustration $r_h^*$ is not a power law: sparse networks accommodate antagonistic ties at the
cost of cooperative ones, which stops paying near $\langle k\rangle\approx10$; the theory reproduces this
without adjustment. In weakly frustrated dense networks the ground state is a near-consensus following the
sign of the aggregate field in $93$--$100\%$ of realisations at small $r_h$, against near-chance agreement in
frustrated ones.
\end{abstract}

\maketitle

\section{Introduction}

Ising models with quenched random fields have been used as models of collective choice since the
first sociophysics papers: Galam, Gefen and Shapir described a strike as a mean-field Ising
transition \cite{galam1982}, Galam treated group decision making as a random-field Ising model (RFIM)
at zero temperature \cite{galam1997}, Brock and Durlauf formulated discrete choice with social
interactions as the same mean-field problem and showed that the ratio of private incentives to
social coupling controls the number of equilibria \cite{brock2001}, and Michard and Bouchaud used
the RFIM to explain abrupt collective opinion shifts \cite{michard2005,bouchaud2013}; see
Refs.~\cite{castellano2009,galam2008,starnini2026} for reviews. Tiwari, Yang and Sen added contrarian and inflexible
agents to a random-field Ising electorate and compared short and long campaigns \cite{tiwari2021}.
Recently, Korbel, Dahdoul and Thurner calibrated an
RFIM with a bimodal field to US House elections and found a polarisation transition, with hysteresis,
above a critical campaign spending \cite{korbel2026}, Bukina and Shepelyansky found a transition from
elite-dominated to crowd-dominated opinion on an Ising network of scientific collaborations
\cite{bukina2026}, and Devauchelle, Szymczak and Nowakowski showed
that an Ising electorate that dislikes the majority opinion settles into near 50--50 splits
\cite{devauchelle2024}. None of these three varies the network's density together with a fraction of
antagonistic ties. In all of these, the organising quantity is the
strength of an individual's own preference (a local field $h$) relative to the social coupling $J$. We call
the first conviction and the second conformity: a voter's conviction is the field $r_hh_i$ pulling them towards
their own preference, and conformity is the pull $\sum_jJ_{ij}s_j$ of their social ties, which a cooperative tie
exerts towards agreement and an antagonistic one towards disagreement.
Disorder in the \emph{sign} of the coupling adds frustration: a fraction of antagonistic ties turns
the model into a spin glass in a random field, the setting of structural-balance theory on signed
networks \cite{harary1953,antal2005,facchetti2011,facchetti2012,gallo2024}. Spin-glass Hamiltonians have also been used to model how homophily within groups and heterophily between them assemble social groups, with a first-order transition between grouped and disordered states \cite{korbel2023}. In opinion dynamics on signed
networks, whether the dynamics stays frustrated depends on edge density \cite{kuroda2023}, and negative
ties and the density of connections between communities affect bi-polarisation non-monotonically
\cite{liu2025}; these are dynamical models and do not address where the ground state passes from
coupling- to field-dominated.

The statistical physics of this model class on sparse random graphs is well developed. Dilute spin
glasses with ferromagnetically biased couplings have ferromagnetic, spin-glass and paramagnetic
phases whose boundaries depend on the graph's branching factor \cite{viana1985,mezard2001,castellani2005};
the Ising model on heterogeneous networks orders at a temperature set by $\langle k^2\rangle/\langle
k\rangle$ rather than $\langle k\rangle$ \cite{dorogovtsev2002,dorogovtsev2008}; the RFIM on Bethe
lattices is solvable at zero temperature \cite{dhar1997}, with disorder-controlled field-driven avalanches \cite{sethna1993}; and in mean-field theory a
discrete field distribution can make the order--disorder transition first order through a
tricritical point \cite{aharony1978,schneider1977}, whose existence on random graphs depends on the
connectivity \cite{doria2015}. The exchange and random-field contributions to the energy have been computed
separately and exactly, as functions of disorder and coordination number, for the metastable field-driven
dynamics of the $T=0$ RFIM on the Bethe lattice \cite{illa2005}; here the same decomposition is applied to the
global ground state with frustrated bonds, and the point where the two contributions are equal is the
observable. Whether the RFIM on sparse graphs has a
spin-glass phase is itself subtle \cite{krzakala2010}, and with Gaussian couplings and fields on random
graphs the joint spin-glass/random-field problem departs from fully connected theory at low connectivity
\cite{erichsen2021}. Without fields, zero-temperature dynamics on random graphs orders only above a mean
degree that grows with $N$ \cite{pournaki2023}.

For a model of an electorate these results leave a quantitative question open: on finite, sparse,
degree-capped social graphs, where does the crossover between coupling-dominated and field-dominated
ground states sit, and how does its location depend on density once part of the social ties are
antagonistic? In the unfrustrated case the answer is energy bookkeeping, since the coupling energy per
voter grows like the number of ties, $\propto\langle k\rangle$. With antagonistic ties the ground state
cannot satisfy every bond, and the coupling energy of a $\pm J$ spin glass grows only like
$\langle k\rangle^{1/2}$ \cite{viana1985,mezard2001,boettcher2003}. We show that the crossover tracks the
coupling energy of the zero-field ground state, to within a nearly constant factor, so that over the
connectivities studied ($\langle k\rangle=3$--$28$) frustration changes its effective density exponent from
about $1$ to about $1/2$, and we measure this change with certified
ground states, across system sizes and graph families, and against zero-temperature theory. The two
energy scalings are established ferromagnetic and spin-glass results, and the change of exponent follows
from them once the crossover is tied to the zero-field coupling energy. The contribution here is
therefore not a new energy law. It is the crossover as an observable of the global ground state, its
connection to those laws, a parameter-free and pre-registered test of zero-temperature cavity theory
against exact ground states, and the localisation of where that theory fails.

\textbf{Relation to Galam's coalition model.} Galam's bimodal-coalition model \cite{galam1996}
already combines individual propensities (local fields) with signed pairwise couplings in an Ising
Hamiltonian and studies the stability of its ground states; it has since been applied to
international alliances \cite{galam2023}. Our Hamiltonian, Eq.~(\ref{eq:hamiltonian}), has the same
form. It differs in four respects. (i) The couplings live on sparse social graphs with up to $720$
voters and realistic degree caps rather than on a small set of fully connected actors, so density and
graph family become variables. (ii) Field magnitudes come from survey-based effect sizes rather than
being set by hand. (iii) Couplings are frustrated by i.i.d.\ sign disorder at a tunable rate $\alpha$
rather than by a fixed alliance structure; the structured, Galam-type case in which signs follow two
camps is the subject of the companion paper \cite{romero2026row1}. (iv) Ground states are computed
exactly and compared instance by instance with zero-temperature cavity theory. ``Contrarian'' also
means something different here. Galam's contrarians are \emph{agents} who adopt the opposite of the
local majority \cite{galam2004}, a property of a site; a negative $J_{ij}$ in Eq.~(\ref{eq:hamiltonian})
is a property of a single \emph{tie}: it lowers the energy when the two voters hold opposite opinions,
whatever they do in their other ties. It encodes a conflicting relationship, a preference for disagreement
on this tie, not necessarily personal hostility. We call such ties antagonistic and $\alpha$ their
fraction.
Agent-level heterogeneity of other kinds, such as inflexible agents \cite{galam2007} or anticonformity
in $q$-voter models \cite{nyczka2012}, is not modelled.

\textbf{Contributions.} (1) A parameter-free ensemble cavity theory, with its predictions recorded before the
exact ground states were computed, tested at seven new densities, together with population mean field and the
cavity method on each graph; its systematic error at intermediate frustration, which coincides with an unstable
replica-symmetric solution in a partly ordered population; and its sign-symmetric branch for $\alpha>1/2$
(Secs.~\ref{sec:theory}--\ref{sec:ensemble}). (2) The field-share crossover $r_h^*$ as an observable, its
approximate proportionality to the zero-field coupling energy inside a rigorous per-instance window of a factor
of two either way (Appendix~\ref{app:bounds}), its closeness to the point at which voters held against their
own field switch, and the resulting change of the effective density exponent from about $1$ to about $1/2$
over $\langle k\rangle=3$--$28$, with its departure from a power law at intermediate frustration
(Secs.~\ref{sec:crossover}--\ref{sec:families}). (3) Certified ground states for 6,750 instances of a
$(m,r_h,\alpha)$ regime map and for the new sweeps up to $N=180$, except part of the densest frustrated
random-regular cells, where the certified fraction is reported and the rest is cross-checked
(Sec.~\ref{sec:exact}). (4) The probability that the collective outcome follows the aggregate field
(Sec.~\ref{sec:outcome}). (5) The exact low-energy landscape and the overlap distribution from parallel
tempering at three sizes (Secs.~\ref{sec:landscape}--\ref{sec:pq}). Several results confirm known physics in
this setting rather than establish it: the $\langle k\rangle$ and $\langle k\rangle^{1/2}$ energy laws and the
approach of the spin-glass prefactor to its Sherrington--Kirkpatrick value, the spin-glass-like antiferromagnet
on non-bipartite graphs, the size of the replica-symmetric energy error, the first-order mean-field transition
for a discrete field distribution, the late alignment of hubs, and percolation of the antagonistic subgraph at
the Molloy--Reed threshold.

\section{Model and Methods}

\subsection{Hamiltonian}

$N$ voters carry spins $s_i=\pm1$ and interact on a graph $G=(V,E)$ through
\begin{equation}
H(\bm s)=-r_h\sum_i h_i s_i-\sum_{(i,j)\in E}J_{ij}s_is_j .
\label{eq:hamiltonian}
\end{equation}
The quenched fields $h_i$ are drawn independently of the graph from the tier distribution of
Table~\ref{tab:params} and multiplied by the swept field--coupling ratio $r_h$. Each edge has a
magnitude $|J_{ij}|$ from the four-tier distribution of Table~\ref{tab:params}
($\langle|J|\rangle=0.31$, $\langle J^2\rangle^{1/2}=0.34$) and sign $-1$ with probability $\alpha$,
independently of everything else. The field distribution is nearly symmetric (mean $-0.04$ against
$\langle|h|\rangle=1.10$, including the $\times1.30$ multiplier; the sign of $h$ labels the two parties), discrete, and has $24\%$ of voters
at or near $h=0$. The model is therefore an RFIM with a trimodal field distribution and $\pm J$ bond
disorder on a sparse graph. We write $M=N^{-1}\sum_is_i$ for the magnetisation (the vote margin) and
label density by the Barab\'asi--Albert attachment parameter $m\in\{2,5,10\}$. Throughout, density means the mean
degree $\langle k\rangle$, the average number of ties per voter; it is not the network-science density
$2|E|/[N(N-1)]$, which falls with $N$ at fixed $\langle k\rangle$.

\subsection{Field and coupling magnitudes}

Field magnitudes come from Bartels' probit estimates of the effect of party-identification strength on
presidential vote choice, averaged over twelve elections \cite{bartels2000}, converted to
percentage-point effects \cite{long1997} and expressed relative to a 25 percentage-point ceiling
\cite{nickerson2008}; tier shares come from ANES 2020 \cite{anes2020}; leaners exceeding weak
partisans reproduces Ref.~\cite{keith1992}. A $\times1.30$ multiplier on a random $10\%$ of voters
represents affective polarisation \cite{iyengar2019}; it is an illustrative perturbation and no result
depends on it. Coupling magnitudes follow a stylised ordering by tie strength
\cite{granovetter1973,gerber2008,bond2012}; they, and the tie-type frequencies, are stylised, not measured
opinion-alignment coefficients or ego-network compositions.
Because $|J|$ is stylised, $r_h=1$ is an order-of-magnitude reference point rather than a calibrated
location of a real electorate, and all results are reported as functions of $r_h$. Taking the ground state as the
collective outcome is, in discrete-choice terms, the limit of infinite intensity of choice ($\beta\to\infty$ in
Ref.~\cite{brock2001}): every voter adopts the opinion that minimises their own energy given the others. We
adopt it because it removes the arbitrary temperature scale; finite temperature enters only through the
parallel-tempering computation of Sec.~\ref{sec:pq}.

\begin{table}[t]
\caption{Model parameters. Field tiers: base value and population share (ANES 2020); ``true
independents'' carry $h\sim U(-0.1,0.1)$ and the ``other/no answer'' category $h=0$ exactly. A random
$10\%$ of voters have $h$ multiplied by $1.30$. Coupling tiers: magnitude and frequency per edge.}
\label{tab:params}
\begin{ruledtabular}
\begin{tabular}{lrr|lrr}
Field tier & $h$ & share & Tie type & $|J|$ & freq.\\
\hline
Strong D/R & $\mp1.729$ & $0.17/0.14$ & Household & $0.55$ & $0.10$\\
Weak D/R & $\mp1.114$ & $0.12/0.12$ & Neighbour & $0.40$ & $0.40$\\
Leaner D/R & $\mp1.256$ & $0.10/0.11$ & Online friend & $0.25$ & $0.30$\\
Independent & $\approx0$ & $0.16$ & Weak tie & $0.10$ & $0.20$\\
Other & $0$ & $0.08$ & & & \\
\end{tabular}
\end{ruledtabular}
\end{table}

\subsection{Graphs}

The reference family is the Barab\'asi--Albert (BA) graph \cite{barabasi1999} with $m\in\{2,5,10\}$,
pruned to maximum degree $20$ by removing, at each over-cap node, non-bridge edges to its lowest-degree
neighbours. At $N=90$ the realised mean degrees are $\langle k\rangle=3.8$, $8.4$ and $12.4$, and
pruning removes $2.7\%$--$30.5\%$ of edges. Four control families are built per realisation from the
same random stream, with the same fields and bond signs: uncapped BA; the configuration model on the
capped-BA degree sequence (self-loops and multi-edges removed); Erd\H{o}s--R\'enyi graphs with the
same number of edges \cite{erdos1959}; and random-regular graphs with $k=\mathrm{round}\langle
k\rangle$. The first two keep degree heterogeneity; the last two remove it at matched density. Pruning
lowers the hub degrees, which reduces degree heterogeneity ($\kappa=\langle k^2\rangle/\langle k\rangle$) and
changes the degree--degree correlations of BA graphs. The configuration-model control keeps the capped degree
sequence but randomises those correlations, and the uncapped control restores the hubs, so the agreement of
the crossover across families (Sec.~\ref{sec:families}) bounds the effect of the cap. Table~\ref{tab:families}
describes the five families in social terms.

\begin{table*}[t]
\caption{The five graph families. All are built per realisation with the same fields and bond signs, and all
but the uncapped BA graph have the same mean degree as the capped BA graph.}
\label{tab:families}
\footnotesize
\begin{ruledtabular}
\begin{tabular}{@{}p{0.13\textwidth}p{0.30\textwidth}p{0.30\textwidth}p{0.21\textwidth}@{}}
Family & Construction & In social terms & Role in the comparison\\
\hline
Capped BA (reference) & Barab\'asi--Albert preferential attachment \cite{barabasi1999}, each over-cap node
pruned to 20 ties by removing ties to its least-connected neighbours & ``Rich get richer'': newcomers befriend
already popular voters, so a few hubs form; no voter keeps more than 20 ties & The model's reference network\\
Uncapped BA & The same attachment process without the cap & The same, with much larger hubs & Restores the
hubs the cap removes\\
Configuration model & Random wiring of the capped-BA degree sequence (self-loops and multi-edges removed) &
Every voter has as many ties as in the capped BA graph, but who knows whom is random & Keeps the degrees;
removes BA's degree--degree correlations\\
Erd\H{o}s--R\'enyi & Uniformly random ties, same number as the capped BA graph \cite{erdos1959} & Random
acquaintance: most voters have close to the average number of ties, and there are no real hubs & Removes
most degree heterogeneity\\
Random-regular & Random wiring in which every node has degree $k=\mathrm{round}\langle k\rangle$ & Every voter
has exactly the same number of ties & Removes degree heterogeneity entirely\\
\end{tabular}
\end{ruledtabular}
\end{table*}

\subsection{Exact ground states}

Ground states are computed as binary quadratic programs with Gurobi \cite{gurobi} at zero relative and
$10^{-9}$ absolute optimality gap, using the optimum at the previous $r_h$ as a warm start; status
OPTIMAL certifies global optimality \cite{hartmann2002}. Every instance with $N\le180$ reported below
is certified, except where a certified fraction is stated (the densest frustrated random-regular graphs of
Secs.~\ref{sec:families} and \ref{sec:ensemble}), where the best incumbent after 120\,s is used. The solver licence available limits this to $N\le180$; for $N=360$ and $720$ we use the
lowest energy of 200 simulated-annealing (SA) reads of 4000 sweeps on a geometric schedule
$\beta\in[0.1,100]$ (dwave-neal \cite{neal}), validated against the certified optima at $N=180$
(Sec.~\ref{sec:exact}). The 20 lowest configurations of an instance are enumerated exactly with
Gurobi's solution pool. Every realisation in every cell has its own seed (NumPy SeedSequence), so no
two cells share a network. Within a cell, a realisation is followed along a 24-point $r_h$ grid,
logarithmic in $[0.1,8]$, so points along one curve are paired and realisations are independent.

\subsection{Observables}
\label{sec:observables}

The field share of the ground-state energy is $f=|E_h|/(|E_h|+|E_J|)$, with $E_h=-r_h\sum_ih_is_i$ and
$E_J=-\sum_{(i,j)}J_{ij}s_is_j$ evaluated on the ground state. The crossover $r_h^*$ is the value at
which $f=1/2$, found for each realisation by linear interpolation in $\ln r_h$ between grid points
(the grid spacing is $0.19$ in $\ln r_h$, well below the realisation-to-realisation spread); its width
is $w=\ln(r_{0.75}/r_{0.25})$, the log-distance between the $f=1/4$ and $f=3/4$ crossings. The absolute
values do not affect $r_h^*$: over all 3,800 realisations in this paper (the 3,440 solved with the exact solver and the 360 annealing runs at $N=180$--$720$ used to validate the heuristic and to extend the sizes), $E_J>0$ occurs only deep in the
field-dominated regime ($f\ge0.95$) and $E_h>0$ only at $r_h\le0.15$ ($f<0.012$), so both terms are negative
wherever $f$ is near $1/2$. $f(r_h)$ is non-decreasing on the grid in all but 38 realisations, whose dips are
at most $0.022$, and every realisation crosses $f=1/2$ exactly once; that crossing defines $r_h^*$. We
also record the magnetisation $M$, the coupling energy per spin of the zero-field ground state,
$e_J(0)=-E_J/N$ of the ground state at $r_h=0$, the pure $\pm J$ ground state of the same graph and
couplings (each realisation solved again at zero field: annealing start, the exact solver with a 120\,s
limit and, where it does not certify, heavy annealing, Sec.~\ref{sec:ensemble}), the number of spins with zero local
field (each implies an exactly degenerate ground state), and whether $\mathrm{sign}\,M=
\mathrm{sign}\sum_ih_i$. Intervals are percentile bootstraps over realisations (2000 resamples) unless
stated otherwise.

\subsection{Zero-temperature theory}

The population mean field at $T=0$ is $M=\mathbb E_h\,\mathrm{sign}(r_hh+J_{\rm eff}M)$, the
zero-temperature limit of the Curie--Weiss and Brock--Durlauf equations \cite{brock2001,schneider1977},
with $J_{\rm eff}=c\,\langle|J|\rangle(1-2\alpha)$ and either $c=\langle k\rangle$ or the Bethe
branching factor $c=\kappa-1$, $\kappa=\langle k^2\rangle/\langle k\rangle$ \cite{dorogovtsev2002}. Its
ground-state energies per spin are $E_h/N=-r_h\mathbb E_h[hs]$ and $E_J/N=-(\langle
k\rangle/2)\langle|J|\rangle(1-2\alpha)M^2$, with $\langle k\rangle/2$ edges per spin. The cavity
method is run on each actual graph: zero-temperature (max-sum) belief propagation with cavity biases
$u_{i\to j}=\mathrm{sign}(J_{ij}h_{i\setminus j})\min(|J_{ij}|,|h_{i\setminus j}|)$ and cavity fields
$h_{i\setminus j}=r_hh_i+\sum_{k\in\partial i\setminus j}u_{k\to i}$, iterated with damping $1/2$ to a
tolerance of $10^{-9}$ (at most 2000 iterations) and decoded as $s_i=\mathrm{sign}(r_hh_i+\sum_{k\in
\partial i}u_{k\to i})$ \cite{mezard2009}. It is exact on trees; on these loopy graphs it is the Bethe
approximation. We record whether it converges and whether its decoded configuration is the exact
ground state.

Neither of these predicts the crossover from the model's distributions alone: the population mean field
ignores the graph, and the cavity method needs each graph. We therefore also solve the replica-symmetric
zero-temperature cavity equations on the configuration-model ensemble of each graph family by population
dynamics \cite{mezard2001}. A population of $10^5$ cavity fields is updated as $H=r_hh+\sum_{l=1}^{d}
u(J_l,H_l)$, with $d$ drawn from the excess-degree distribution $(d+1)P(d+1)/\langle k\rangle$, $h$ from
the field distribution of Table~\ref{tab:params}, $J_l$ from the coupling distribution with sign $-1$ at
rate $\alpha$, and the $H_l$ from the population. The ground-state energy per spin is
$E/N=\langle f_{\rm site}\rangle-(\langle k\rangle/2)\langle f_{\rm link}\rangle$ with
$f_{\rm site}=-\sum_{l=1}^{k}\max(|H_l|,|J_l|)-|r_hh+\sum_{l=1}^{k}u(J_l,H_l)|$ for a site of degree
$k\sim P(k)$ and $f_{\rm link}=-\max_{s,s'}(H_1s+H_2s'+Jss')$ \cite{mezard2009} (terms $\pm\sum_l|H_l|$
that cancel between the site and link contributions are omitted from both); the field term is
$-r_h\langle hs\rangle$, $s$ being the sign of the full local field. The only inputs are $P(k)$, pooled
from 300 generated $N=90$ graphs of the family and density in question, and the distributions of $h$ and
$J$: no instance is solved and nothing is fitted. The population is initialised with independent
standard normal fields and equilibrated for 150 sweeps at $r_h=10^{-6}$; $r_h$ is then raised through 33
values in $[0.1,8]$, the population carried over, with 30 equilibration sweeps and 8 measurements each
(a sweep updates every member once). The Monte Carlo error of the resulting $r_h^*$ is estimated from four
independent seeds and a fourfold population, and its dependence on the protocol from longer runs, a
descending $r_h$ sweep and an ordered initial population (Sec.~\ref{sec:ensemble}). The
solution is replica symmetric. For a $\pm J$ spin glass at $T=0$ replica-symmetry breaking changes the
ground-state energy by less than $1\%$ at low connectivity \cite{mezard2003,boettcher2003} and by $4.6\%$
in the fully connected limit \cite{parisi1980}, which is why a replica-symmetric energy balance can be
accurate at $\alpha=1/2$ although the true ground state is glassy. For $\alpha>1/2$
we restrict the population to its sign-symmetric branch, by multiplying every field by an independent
random sign after each sweep (Sec.~\ref{sec:ensemble}). In the ferromagnetic limit the energy is
$-(\langle k\rangle/2)\langle|J|\rangle$ to $0.2\%$, and the field term equals $r_h\,dE/dr_h$ to within
$1\%$, as it must at $T=0$. All theory in this paper is at $T=0$, the temperature of the ground-state
observable; we use no finite-temperature mean-field bifurcations, whose locations depend on $\beta$.

\subsection{Parallel tempering}

The overlap distribution is computed with parallel tempering \cite{hukushima1996} on $m=10$ graphs: 16
temperatures geometric in $\beta\in[0.1,2]$, single-spin Metropolis sweeps, a swap attempt between
neighbouring temperatures after every sweep, and two independent chains per instance, so that the
overlap $q=N^{-1}\sum_is_i^{(1)}s_i^{(2)}$ is measured between chains at equal temperature. Runs last
$2^{14}$ sweeps ($N=45$, $90$) or $2^{15}$ sweeps ($N=180$). Observables are averaged in logarithmic
bins $[2^{k-1},2^k)$, and a cell is accepted as equilibrated when the disorder averages of $\langle
q^2\rangle$ and of the energy in the last two bins agree within errors \cite{katzgraber2001}. We also
record swap acceptance and the number of hottest-to-coldest traversals of the temperature ladder.

\section{Results}

\subsection{Exact ground states and the annealer}
\label{sec:exact}

We first certified the ground states of the regime map used in earlier work on this model: three
densities, nine $r_h$ values ($0.25$--$16$), five $\alpha$ values and 50 independent realisations per
cell, 6,750 instances at $N=90$, each previously solved by SA with 1000 reads of 1500 sweeps. All 6,750
are certified optimal. SA with its native schedule ($\beta\in[0.1,10]$ at every $r_h$) found the exact
ground state in all 6,750. A second schedule convention, in which the $\beta$ range is divided by $r_h$
to keep the final temperature fixed relative to the field (equivalent to scaling $J$ down rather than
$h$ up; Appendix~\ref{app:schedule}), found it in $6{,}414$ ($95.0\%$); every miss is at $r_h\ge8$, where
this schedule ends too hot for the zero-field voters, and the largest relative energy error is
$1.6\times10^{-3}$. The regime-map statistics are therefore exact; the certification status of every
other data set is stated where it is reported. Exact degeneracy is
common at low density: $31\%$, $10\%$ and $7\%$ of instances at $m=2$, $5$, $10$ have at least one spin
with zero local field in the ground state. At $N=180$, the heuristic used for larger systems reached the
certified optimum in $2{,}877$ of $2{,}880$ $(\text{realisation},r_h)$ points (120 realisations
$\times$ 24 values of $r_h$); the three misses have relative energy errors of at most
$3.8\times10^{-3}$. Appendix~\ref{app:sa} reports SA success against the exact
optimum for 18 combinations of final $\beta$, sweeps and reads.

\subsection{The field--coupling crossover and its density scaling}
\label{sec:crossover}

\begin{figure*}[t]
\centering
\includegraphics[width=\textwidth]{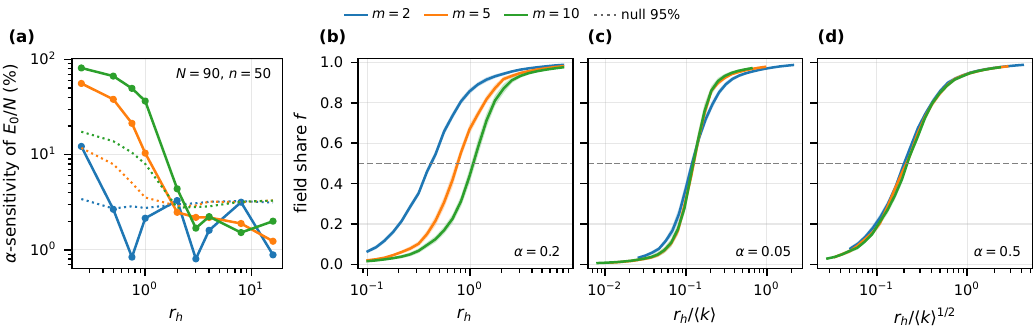}
\caption{Field--coupling crossover from exact ground states. (a) $\alpha$-sensitivity of $E_0/N$ (spread
of cell means across $\alpha\in\{0.05,0.2,0.5,0.8,1\}$, as a percentage of their mean) against $r_h$, by
density; dotted, 95th percentile of the permutation null. $N=90$, 50 realisations per cell. (b) Field
share $f$ of the ground-state energy at $\alpha=0.2$, $N=90$ (30 realisations; bands, 95\% CI of the
mean). (c) The same at $\alpha=0.05$ against $r_h/\langle k\rangle$. (d) The same at $\alpha=0.5$ against
$r_h/\langle k\rangle^{1/2}$. Dashed, $f=1/2$.}
\label{fig:regime}
\end{figure*}

Figure~\ref{fig:regime}(a) shows the sensitivity of the exact ground-state energy to $\alpha$, the spread
of the cell-mean $E_0/N$ across the five $\alpha$ values as a percentage of its mean, against a
permutation null ($\alpha$ labels shuffled within each $(m,r_h)$ slice, 2000 permutations; dotted). It
falls from $81\%$, $56\%$ and $12\%$ ($m=10$, $5$, $2$) at $r_h=0.25$ to the null level at large $r_h$. An
intrinsic measure of where this happens, the $r_h$ at which the spread on the fine grid falls to half its
$r_h=0.1$ value, is $0.28$, $0.59$ and $0.83$ for $m=2$, $5$, $10$ at $N=90$, and $0.28$, $0.64$, $0.88$
at $N=45$. In the field-dominated regime the ground state is set by $\mathrm{sign}(h_i)$ whatever the
bond signs, so the decay itself is expected; its location is the informative quantity.

\begin{figure*}[t]
\centering
\includegraphics[width=\textwidth]{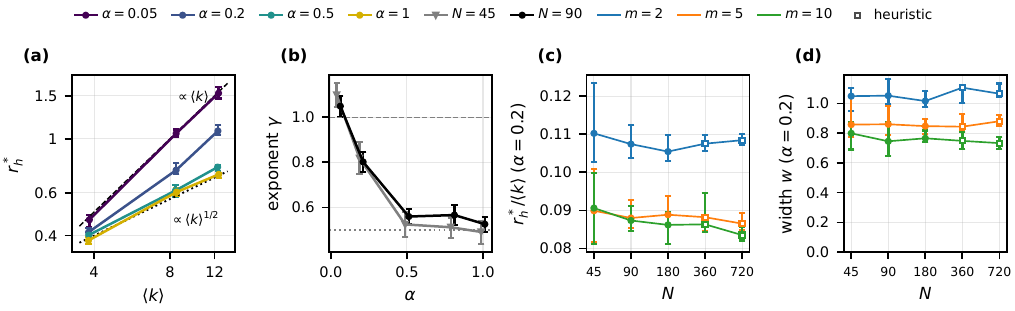}
\caption{Density scaling and size dependence of the crossover. (a) $r_h^*$ (median over 30 realisations,
95\% CI) against $\langle k\rangle$ at $N=90$; dashed and dotted guides $\propto\langle k\rangle$ and
$\propto\langle k\rangle^{1/2}$. (b) Exponent $\gamma$ of $r_h^*\propto\langle k\rangle^\gamma$, fitted over
all realisations of the three densities, with bootstrap 95\% CIs. (c) $r_h^*/\langle k\rangle$ and (d)
crossover width $w$ at $\alpha=0.2$ against $N$; filled, certified exact; open, heuristic ground states.}
\label{fig:crossover}
\end{figure*}

The field share $f(r_h)$ [Fig.~\ref{fig:regime}(b)] rises smoothly from near zero to near one, and its
half-point $r_h^*$ moves to larger $r_h$ with density. Table~\ref{tab:rstar} gives $r_h^*$ for every
$(m,\alpha)$ at $N=90$. At $\alpha=0.05$ the three densities collapse onto one curve when $r_h$ is divided
by $\langle k\rangle$ [Fig.~\ref{fig:regime}(c)]; at $\alpha=0.5$ they collapse when it is divided by
$\langle k\rangle^{1/2}$ [Fig.~\ref{fig:regime}(d)]. Fitting $\ln r_h^*=\gamma\ln\langle k\rangle+c$ over
all realisations of the three densities [Fig.~\ref{fig:crossover}(a,b)] gives $\gamma=1.05$
$[1.00,1.10]$ at $\alpha=0.05$, $0.80$ $[0.76,0.85]$ at $\alpha=0.2$, and $0.56$ $[0.52,0.60]$, $0.57$
$[0.52,0.61]$ and $0.53$ $[0.49,0.56]$ at $\alpha=0.5$, $0.8$ and $1$ ($N=90$; $N=45$ agrees within
errors). Sec.~\ref{sec:ensemble} extends these fits to seven capped-BA densities and to random-regular
graphs with $\langle k\rangle=3$--$28$.

\begin{table}[t]
\caption{Crossover $r_h^*$ (field share $1/2$) at $N=90$: crossing of the realisation-averaged curve, with
a bootstrap 95\% CI that includes interpolation. 30 certified realisations per cell;
$\langle k\rangle=3.8$, $8.4$ and $12.4$ for $m=2$, $5$, $10$.}
\label{tab:rstar}
\footnotesize
\begin{ruledtabular}
\begin{tabular}{@{}lccc@{}}
$\alpha$ & $m=2$ & $m=5$ & $m=10$\\
\hline
0.05 & 0.455 [0.437, 0.474] & 1.05 [1.02, 1.08] & 1.53 [1.46, 1.59]\\
0.2 & 0.415 [0.400, 0.432] & 0.754 [0.727, 0.778] & 1.07 [1.02, 1.12]\\
0.5 & 0.392 [0.377, 0.404] & 0.615 [0.590, 0.642] & 0.763 [0.744, 0.781]\\
0.8 & 0.389 [0.374, 0.403] & 0.615 [0.589, 0.643] & 0.748 [0.725, 0.776]\\
1.0 & 0.379 [0.370, 0.391] & 0.601 [0.587, 0.617] & 0.707 [0.684, 0.729]\\
\end{tabular}
\end{ruledtabular}
\end{table}

The exponent follows from where the crossover sits. Per realisation, $r_h^*$ is close to the coupling
energy per spin of the zero-field ground state divided by the mean field magnitude: the ratio
$r_h^*/e_J(0)$ has median $0.91$ (interquartile range $0.85$--$0.98$) over all 1,080 realisations, and
its cell medians lie between $0.85$ and $1.00$ in all 36 $(N,m,\alpha)$ cells, while $r_h^*$ itself
varies fourfold, and $1/\langle|h|\rangle=0.90$. The crossover is therefore where the largest available
field energy, $r_h\langle|h|\rangle$ per spin, matches the coupling energy the zero-field ground state
has. This is an empirical proportionality for our field distribution, not an identity: the argument
fixes the prefactor only through $1/\langle|h|\rangle$, and a different field distribution would change it.
The proportionality itself, up to a factor of two either way, is not empirical, given that both energy terms
are negative at the crossover (verified in every realisation, Sec.~\ref{sec:observables}). Comparing the ground state
at the crossover with the zero-field ground state and with a field-aligned state bounds $r_h^*$ in every
instance (Appendix~\ref{app:bounds}),
\begin{equation}
\frac12\;\le\;\frac{r_h^*\,\langle|h|\rangle}{e_J(0)}\;\le\;2+\frac{\varepsilon_h}{e_J(0)},
\label{eq:window}
\end{equation}
where $\langle|h|\rangle$ is the realisation's own mean field magnitude and $\varepsilon_h$ is the coupling
energy per spin of a field-aligned configuration, zero on average for fields drawn independently of the
graph. All 3,440 realisations solved with the exact solver (210 of them not certified at $r_h=0$, Sec.~\ref{sec:ensemble}; the 360 annealing runs at $N=180$--$720$ have no $e_J(0)$) lie inside this window, at a median of $1.00$ (range
$0.73$--$1.53$; the upper limit is $1.47$--$2.44$). The data sit at its geometric centre; that position, not
the window, is the empirical part.
The energy split also marks where voters change their vote. Take the partisans and leaners ($|h_i|\ge1$) whose
ground-state opinion at the smallest $r_h$ is against their own field, about 30 per realisation: half of them
have switched to their field for good at $0.93\,r_h^*$ (median over the 900 realisations of
Sec.~\ref{sec:families} with $\alpha\le0.2$; interquartile range $0.81$--$0.98$). Cell medians are
$0.92$--$0.99$ at $m=10$ and $0.75$--$0.91$ at $m=2$, where sparse networks release their
dissenters somewhat earlier. In dense networks $r_h^*$ is thus close to the point at which the network stops
holding the typical dissenting partisan, and in sparse ones somewhat above it (Appendix~\ref{app:switching}); we
measured this for $\alpha\le0.2$ only.
The window is energy bookkeeping. What is not bookkeeping is how $e_J(0)$ depends on density. At
$\alpha=0.05$ it equals $(0.90$--$0.92)\,(\langle k\rangle/2)\langle|J|\rangle$ at every density and size (the
ground state satisfies $94$--$95\%$ of bonds), so it grows like $\langle k\rangle$. At $\alpha\ge0.5$ the ground
state satisfies only $64$--$82\%$ of bonds, decreasing with density, and $e_J(0)=(0.61$--$0.70)\,
(\langle k\rangle\langle J^2\rangle)^{1/2}$ at every density and size of the main sweeps ($\langle k\rangle\le13$), the $\langle k\rangle^{1/2}$
dependence of the ground-state energy of a $\pm J$ spin glass of finite connectivity, whose prefactor
approaches the Sherrington--Kirkpatrick value $0.763$ \cite{parisi1980,viana1985,mezard2001,boettcher2003} from below
($0.73$ on random-regular graphs with $k=20$ and with $k=28$ at $\alpha=1/2$ and $N=90$; $0.73$ and $0.74$, respectively, at $N=180$; Sec.~\ref{sec:ensemble}).
Both energy laws are rigorous, so Eq.~(\ref{eq:window}) turns them into statements about $r_h^*$. For
$\alpha<1/2$, $\sum_{(i,j)}J_{ij}/N\le e_J(0)\le\sum_{(i,j)}|J_{ij}|/N$ (the all-aligned state and the
all-satisfied bound), so $e_J(0)$, and with it $r_h^*$, grows linearly in $\langle k\rangle$ on any graph. For
unit couplings on Erd\H{o}s--R\'enyi and random-regular graphs, the ground-state energy of the $\pm J$ spin
glass ($\alpha=1/2$) and of the antiferromagnet ($\alpha=1$) is $P_*\sqrt{\langle k\rangle}\,[1+o(1)]$ per spin
at large degree, with $P_*\approx0.763$ the Sherrington--Kirkpatrick ground-state energy
\cite{dembo2017,sen2018}; there $r_h^*$ therefore grows like $\langle k\rangle^{1/2}$. The bounds fix the
exponents, not the prefactor, and the theorems are asymptotic in $\langle k\rangle$; they do not cover the
capped BA graphs or the four-tier couplings, for which the exponents of this section remain measured ones.
Antagonistic ties thus change the density scaling of the field--coupling crossover from $\langle
k\rangle$ to $\langle k\rangle^{1/2}$: in a frustrated network, adding ties adds coupling energy only
incoherently. At $\alpha=1$ the bonds are all antiferromagnetic, but on these non-bipartite graphs they
are frustrated by odd cycles and behave like the $\alpha=0.5$ spin glass (Sec.~\ref{sec:ensemble}).

\textbf{Size dependence.} Between $N=45$ and $180$ (certified) the density-normalised crossover is
unchanged within errors [Fig.~\ref{fig:crossover}(c); $r_h^*$ itself drifts down slightly at $m=10$
because the degree cap removes a larger fraction of edges at larger $N$, lowering $\langle k\rangle$
from $13.2$ to $11.4$]. The width of the crossover in a single realisation, $w\approx0.75$--$1.05$ in
$\ln r_h$, also does not shrink [Fig.~\ref{fig:crossover}(d)]; with the validated heuristic the same holds to $N=720$ at
$\alpha=0.05$ and $0.2$ ($w=0.56$--$1.06$, $r_h^*/\langle k\rangle=0.120$--$0.131$ at $\alpha=0.05$ and
$0.084$--$0.108$ at $\alpha=0.2$ for $N=180$--$720$), and the density exponent at $N=720$ is
$\gamma=1.02$ ($\alpha=0.05$) and $0.77$ ($\alpha=0.2$), against $1.07$ and $0.80$ at $N=180$. This does not
rule out a phase transition. Because the fields have nearly zero mean, raising $r_h$ raises the field
disorder relative to $J$, and the $T=0$ random-field Ising model on sufficiently homogeneous random graphs
(scale-free ones with degree exponent above three) has a disorder-driven transition from an ordered to a
disordered ground state, continuous or discontinuous depending on the field distribution \cite{lee2006}. The field share is an energy ratio, which can remain smooth across a
transition in $M$, so its width does not test for one; we do not locate such a transition, and four certified
sizes and a heuristic extension to $N=720$ could not settle its existence or order.

\subsection{Graph family}
\label{sec:families}

\begin{figure}[t]
\centering
\includegraphics[width=\columnwidth]{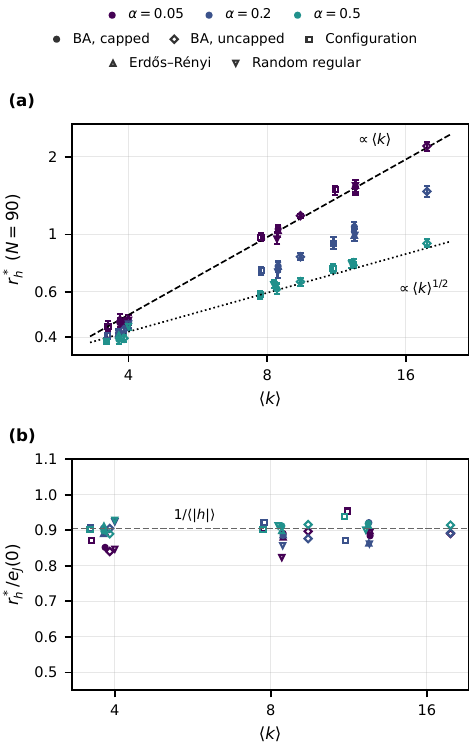}
\caption{Crossover by graph family at $N=90$ (30 realisations per point). (a) $r_h^*$ of the mean
field-share curve (95\% bootstrap CIs) against $\langle k\rangle$ for $m=2,5,10$; colour, $\alpha$; marker,
family (filled, capped BA). Guides $\propto\langle k\rangle$ and $\propto\langle k\rangle^{1/2}$. (b)
$r_h^*/e_J(0)$; the dashed line is $1/\langle|h|\rangle$. Families share the fields and bond-sign draws.}
\label{fig:families}
\end{figure}

\begin{figure}[t]
\centering
\includegraphics[width=\columnwidth]{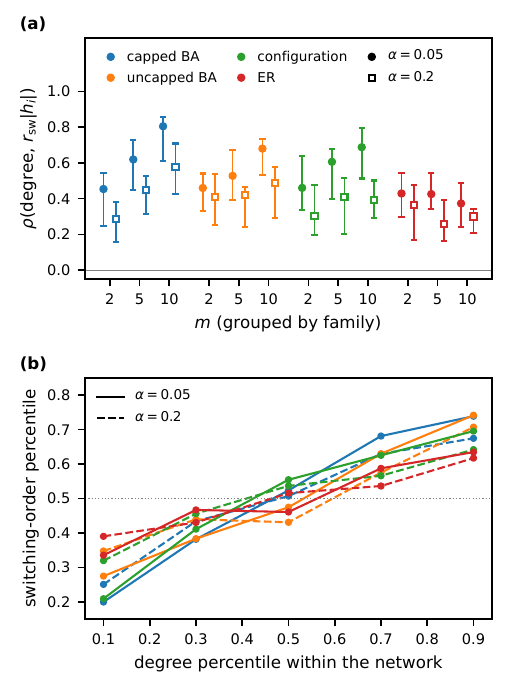}
\caption{Hubs follow their own fields last (exact ground states, $N=90$, 30 realisations per cell; voters
with $|h_i|\ge1$ whose ground-state opinion at the smallest $r_h$ is against their own field). (a) Median
Spearman $\rho$ between degree and the field energy $r_{\rm sw}|h_i|$ at which a voter switches to its field
for good, by family and $m$ (bars, interquartile range over realisations); filled, $\alpha=0.05$; open,
$\alpha=0.2$. (b) Mean switching-order percentile against degree percentile within each network, $m=10$;
dotted, no dependence on degree. Random-regular graphs, with one degree, are not shown.}
\label{fig:hubs}
\end{figure}

Neither the degree cap nor degree heterogeneity is responsible for the density scaling. We rebuilt every
$N=90$ realisation at $\alpha\in\{0.05,0.2,0.5\}$ on four control families with the same fields and
bond-sign draws: uncapped BA, the configuration model on the capped-BA degree sequence (self-loops and
multi-edges removed), Erd\H{o}s--R\'enyi graphs with the same number of edges, and random-regular graphs
with $k=\mathrm{round}\langle k\rangle$ (30 realisations per cell, all ground states certified except
$1$--$14\%$ of the (realisation, $r_h$) points in the $m=10$, $\alpha=0.5$ cells of the uncapped BA, ER and
random-regular families, which hit the 120\,s limit and use the best incumbent). At matched $\langle k\rangle$ the crossover is the same on every
family [Fig.~\ref{fig:families}(a)]: at $m=10$, $\alpha=0.05$, $r_h^*=1.53$, $1.55$ and $1.55$ for capped BA,
ER and random-regular graphs, and the uncapped BA, whose $\langle k\rangle$ is $43\%$ higher, lies on the
same line. The exponent is $\gamma=1.01$--$1.07$ at $\alpha=0.05$, $0.69$--$0.81$ at $\alpha=0.2$ and
$0.50$--$0.57$ at $\alpha=0.5$ across the five families (fit to the mean-curve crossings; 95\% CIs
$\pm0.04$--$0.06$; at $\alpha=0.2$ the spread is partly one of density range, since $r_h^*$ is not a power
law there and the uncapped BA reaches higher $\langle k\rangle$, Sec.~\ref{sec:ensemble}), and $r_h^*/e_J(0)=0.82$--$0.95$ in all 45 cells [Fig.~\ref{fig:families}(b)]. Degree
heterogeneity affects only the width: at $\alpha=0.05$ the crossover is sharper on homogeneous graphs
($w=0.34$ random-regular, $0.40$ ER, $0.62$ capped BA at $m=10$), because hubs, whose coupling energy is
largest, are the last spins to follow their fields. Among the partisans and leaners whose ground-state opinion
starts against their own field ($\approx30$ per realisation), degree and the field energy $r_h|h_i|$ at which
they switch for good are positively rank-correlated in $97\%$ of the 720 realisations of the four
heterogeneous families at $\alpha\in\{0.05,0.2\}$ (median Spearman $\rho=0.44$; cell medians $0.26$--$0.81$,
largest on BA graphs and smallest on ER graphs), and the top degree quartile is over-represented
$1.4$--$2.5\times$ among the last quarter to switch [Fig.~\ref{fig:hubs}; exact ground states, re-solved on
the same realisations].
This is what heterogeneous mean-field theory leads one to expect: a spin's coupling field grows with its
degree \cite{dorogovtsev2002,dorogovtsev2008}, and hubs hold the random-field Ising model on scale-free
networks in its ordered state \cite{lee2006}. What the check adds is that the ordering holds voter by voter
in exact ground states with antagonistic ties. The effect is strongest on capped BA graphs at $m=10$ (median
$\rho=0.81$ against $0.69$ on the configuration model with the same degrees and the same fields and bond signs;
larger in 24 of 30 paired realisations, Wilcoxon $p<10^{-3}$), where the cap, which removes an over-cap
node's ties to its least-connected neighbours, leaves hubs tied to hubs (degree assortativity $0.34$ against
$-0.04$); at $m=2$ and $5$, where the cap removes few ties, the two families agree (Appendix~\ref{app:wiring}).

\subsection{Zero-temperature theory}
\label{sec:theory}

\begin{figure*}[t]
\centering
\includegraphics[width=\textwidth]{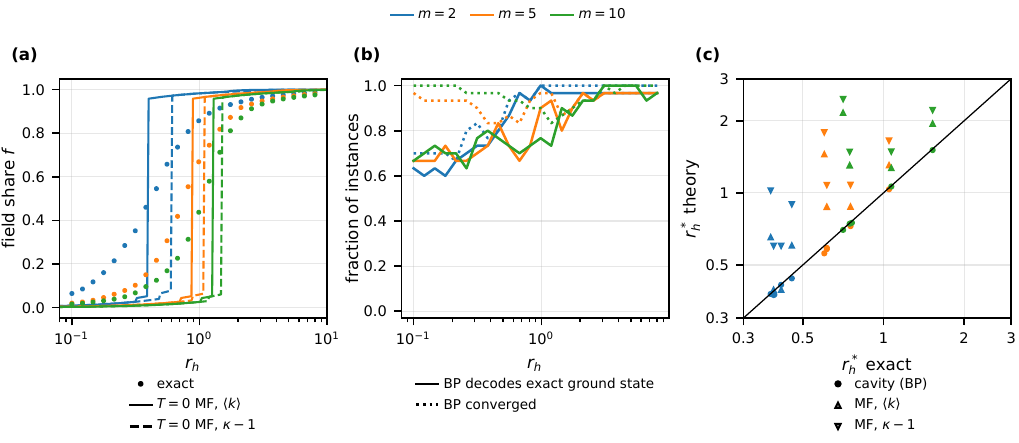}
\caption{Zero-temperature theory against exact ground states ($N=90$). (a) Field share at $\alpha=0.2$:
exact (points) and population mean field with $c=\langle k\rangle$ (solid) and $c=\kappa-1$ (dashed). The
mean-field curves are self-consistent $T=0$ solutions on the ordered branch, not fits; their jump is the
collapse of that branch (Sec.~\ref{sec:theory}).
(b) Fraction of instances in which cavity belief propagation converges (dotted) and decodes the exact
ground state (solid), $\alpha=0.2$. (c) $r_h^*$ from theory against exact, all $(m,\alpha)$ cells: cavity
(filled circles) and population mean field (open triangles; $\alpha<0.5$ only).}
\label{fig:theory}
\end{figure*}

Figure~\ref{fig:theory} compares two levels of zero-temperature theory with the exact ground states. The
population mean field has an ordered branch for $\alpha<1/2$ that disappears discontinuously at a fold,
so its field share jumps rather than rising smoothly [Fig.~\ref{fig:theory}(a)]; with a field
distribution this discrete, the mean-field order--disorder transition is first order
\cite{aharony1978}. The jump is a cascade. On the ordered branch every voter feels the same social pull
$J_{\rm eff}M$; apart from the few voters with the $\times1.30$ multiplier, the minority party's strong
partisans ($|h|=1.73$) defect first, at $r_h\approx
J_{\rm eff}/1.73$ ($0.41$, $0.90$ and $1.33$ for $m=2$, $5$, $10$ at $\alpha=0.2$, where the curves jump),
which lowers $M$ and with it the pull on the weaker partisans, who then defect in turn. The exact curves show
no such jump at any $N$ studied: on a graph each voter's pull depends on their own degree and neighbours, which
spreads the individual thresholds, so that, with field disorder this broad compared with $J$, the network
turns the all-or-nothing collapse of mean field into a gradual crossover. A graph does not remove jumps in
general: on Bethe lattices of coordination four or more the $T=0$ random-field Ising model has a
discontinuity when the disorder is weak \cite{dhar1997}. With $c=\langle k\rangle$ the
mean field places $r_h^*$ within $5$--$33\%$ of the exact value for $\alpha\le0.2$ (ratios $0.95$--$1.33$);
with the Bethe factor $c=\kappa-1$ it overshoots by $1.4$--$2.0\times$, because on these capped graphs
$\kappa-1$ exceeds $\langle k\rangle$ ($\kappa-1=5.8$, $10.5$ and $14.3$ against $\langle k\rangle=3.8$, $8.4$ and $12.4$ at $N=90$) while the ground-state coupling energy does not. For
$\alpha\ge1/2$, $J_{\rm eff}\le0$ and the population theory has no coupling energy to balance against
the field, although the exact $e_J(0)$ is large: it cannot represent frustrated coupling energy at all.

The cavity method on the actual graphs does. Its decoded configuration is the exact ground state in
$82$--$88\%$ of instances for $\alpha\le0.2$ (averaged over the $r_h$ grid) and in $42$--$78\%$ for
$\alpha\ge0.5$, where it often fails to converge at small $r_h$ [Fig.~\ref{fig:theory}(b)]. Its field-share
crossover nonetheless matches the exact one closely in every cell [Fig.~\ref{fig:theory}(c)]: the median
cavity $r_h^*$ is within $8\%$ of the exact median in all 15 $(m,\alpha)$ cells at $N=90$ and within $3\%$ in
10 of them, including frustrated ones (it is never larger than the exact value), because the energy split is insensitive to the few spins on which the
decoding errs.

\subsection{Parameter-free prediction and an out-of-sample test}
\label{sec:ensemble}

\begin{table*}[t]
\caption{The out-of-sample test: 21 cells at densities not used elsewhere in the paper ($N=90$, 30 realisations
each). Exact $r_h^*$ with bootstrap 95\% CI; predicted $r_h^*$ recorded before the exact run, $\pm$ the standard
deviation over four population-dynamics seeds (a Monte Carlo spread, not a confidence interval, so the two
uncertainties are not directly comparable); error of the recorded prediction (a miss is $|\mathrm{error}|>10\%$); $e_J(0)$ exact and predicted;
$r_h^*/e_J(0)$, the median of the per-realisation ratios (not the ratio of the two columns); fraction of certified points over
the curve and at $r_h=0$ (where $e_J(0)$ is measured).}
\label{tab:ensemble}
\footnotesize
\begin{ruledtabular}
\begin{tabular}{@{}llcccccccc@{}}
Graph & $\alpha$ & $\langle k\rangle$ & exact $r_h^*$ & predicted $r_h^*$ & error & $e_J(0)$ exact & $e_J(0)$ pred. & $r_h^*/e_J(0)$ & certified \\
\hline
BA $m=3$ & 0.05 & 5.6 & 0.681 [0.647, 0.714] & 0.669 $\pm$ 0.004 & $-1.86\%$ & 0.783 & 0.768 & 0.87 & 100\%/100\% \\
BA $m=3$ & 0.2 & 5.5 & 0.517 [0.497, 0.538] & 0.504 $\pm$ 0.002 & $-2.64\%$ & 0.593 & 0.593 & 0.85 & 100\%/100\% \\
BA $m=3$ & 0.5 & 5.5 & 0.491 [0.472, 0.509] & 0.495 $\pm$ 0.002 & $+0.79\%$ & 0.542 & 0.563 & 0.91 & 100\%/100\% \\
BA $m=4$ & 0.05 & 7.0 & 0.890 [0.852, 0.929] & 0.834 $\pm$ 0.023 & $-6.20\%$ & 0.981 & 0.970 & 0.92 & 100\%/100\% \\
BA $m=4$ & 0.2 & 7.1 & 0.642 [0.617, 0.667] & 0.604 $\pm$ 0.003 & $-5.96\%$ & 0.735 & 0.717 & 0.87 & 100\%/100\% \\
BA $m=4$ & 0.5 & 7.1 & 0.542 [0.525, 0.564] & 0.568 $\pm$ 0.002 & $+4.80\%$ & 0.620 & 0.647 & 0.88 & 100\%/100\% \\
BA $m=7$ & 0.05 & 10.6 & 1.320 [1.280, 1.368] & 1.361 $\pm$ 0.040 & $+3.15\%$ & 1.488 & 1.450 & 0.88 & 100\%/100\% \\
BA $m=7$ & 0.2 & 10.5 & 0.945 [0.913, 0.981] & 0.826 $\pm$ 0.001 & $-12.54\%$ & 1.026 & 1.003 & 0.93 & 100\%/100\% \\
BA $m=7$ & 0.5 & 10.6 & 0.690 [0.670, 0.710] & 0.710 $\pm$ 0.002 & $+2.91\%$ & 0.760 & 0.818 & 0.91 & 100\%/100\% \\
BA $m=15$ & 0.05 & 12.8 & 1.744 [1.660, 1.836] & 1.616 $\pm$ 0.081 & $-7.36\%$ & 1.788 & 1.836 & 1.03 & 100\%/100\% \\
BA $m=15$ & 0.2 & 12.7 & 1.163 [1.111, 1.230] & 1.046 $\pm$ 0.024 & $-10.03\%$ & 1.212 & 1.212 & 0.98 & 100\%/100\% \\
BA $m=15$ & 0.5 & 12.8 & 0.762 [0.732, 0.795] & 0.787 $\pm$ 0.004 & $+3.25\%$ & 0.839 & 0.889 & 0.90 & 100\%/100\% \\
RR $k=3$ & 0.05 & 3.0 & 0.382 [0.371, 0.392] & 0.371 $\pm$ 0.001 & $-2.74\%$ & 0.426 & 0.430 & 0.90 & 100\%/100\% \\
RR $k=3$ & 0.2 & 3.0 & 0.375 [0.364, 0.387] & 0.371 $\pm$ 0.000 & $-1.05\%$ & 0.409 & 0.413 & 0.92 & 100\%/100\% \\
RR $k=3$ & 0.5 & 3.0 & 0.368 [0.358, 0.379] & 0.370 $\pm$ 0.001 & $+0.66\%$ & 0.406 & 0.414 & 0.90 & 100\%/100\% \\
RR $k=20$ & 0.05 & 20.0 & 2.418 [2.367, 2.494] & 2.498 $\pm$ 0.005 & $+3.29\%$ & 2.810 & 2.788 & 0.85 & 100\%/100\% \\
RR $k=20$ & 0.2 & 20.0 & 1.626 [1.568, 1.681] & 1.466 $\pm$ 0.030 & $-9.83\%$ & 1.890 & 1.868 & 0.85 & 100\%/100\% \\
RR $k=20$ & 0.5 & 20.0 & 1.033 [1.001, 1.067] & 1.031 $\pm$ 0.005 & $-0.25\%$ & 1.106 & 1.166 & 0.95 & 55\%/0\% \\
RR $k=28$ & 0.05 & 28.0 & 3.430 [3.343, 3.519] & 3.717 $\pm$ 0.238 & $+8.37\%$ & 3.907 & 3.903 & 0.88 & 100\%/100\% \\
RR $k=28$ & 0.2 & 28.0 & 2.299 [2.190, 2.400] & 2.183 $\pm$ 0.011 & $-5.04\%$ & 2.589 & 2.577 & 0.89 & 100\%/100\% \\
RR $k=28$ & 0.5 & 28.0 & 1.194 [1.155, 1.232] & 1.225 $\pm$ 0.002 & $+2.66\%$ & 1.308 & 1.392 & 0.91 & 44\%/0\% \\
\end{tabular}
\end{ruledtabular}
\end{table*}

\begin{table}[tbp]
\caption{Density exponent $\gamma$ of $r_h^*\propto\langle k\rangle^\gamma$ at $N=90$: exact, fitted to the
per-realisation crossings (bootstrap 95\% CI) and to the crossings of the mean curves (the definition used
for the prediction), and predicted by the ensemble theory (mean over four seeds). Capped BA, seven densities ($\langle k\rangle=3.8$--$12.8$);
random-regular (RR), $k=3$--$28$ (three densities, $k=3,20,28$, at $\alpha\ge0.8$ and for the theory).
$^\dagger$Sign-symmetric branch, computed after the exact results (Sec.~\ref{sec:ensemble}).}
\label{tab:gamma}
\footnotesize
\begin{ruledtabular}
\begin{tabular}{@{}llccc@{}}
Graph & $\alpha$ & exact (per realisation) & exact (mean curve) & theory \\
\hline
BA & 0.05 & 1.09 [1.05, 1.13] & 1.06 & 1.06 \\
BA & 0.2 & 0.87 [0.83, 0.91] & 0.86 & 0.80 \\
BA & 0.5 & 0.55 [0.52, 0.58] & 0.55 & 0.56 \\
BA & 0.8 & 0.55 [0.52, 0.58] & 0.54 & 0.56$^\dagger$ \\
BA & 1 & 0.52 [0.49, 0.54] & 0.52 & 0.56$^\dagger$ \\
RR & 0.05 & 1.01 [0.99, 1.03] & 1.00 & 1.01 \\
RR & 0.2 & 0.81 [0.79, 0.83] & 0.80 & 0.77 \\
RR & 0.5 & 0.53 [0.51, 0.54] & 0.53 & 0.54 \\
RR & 0.8 & 0.50 [0.49, 0.52] & 0.51 & 0.54$^\dagger$ \\
RR & 1 & 0.50 [0.48, 0.52] & 0.50 & 0.54$^\dagger$ \\
\end{tabular}
\end{ruledtabular}
\end{table}

The ensemble cavity theory has seen no instance. At the three densities of Table~\ref{tab:rstar} it gives
$r_h^*$ within $5\%$ of the exact value in eight of the nine cells with $\alpha\le1/2$ and within $13\%$ in
the ninth ($m=5$, $\alpha=0.2$), $e_J(0)$ within $7\%$, and exponents $\gamma=1.04$, $0.78$, $0.57$
against the exact $1.05$, $0.80$, $0.56$. To test it out of sample, we computed its predictions for
seven densities not used anywhere else in this paper: capped BA with $m=3$, $4$, $7$, $15$
($\langle k\rangle=5.5$--$12.8$) and random-regular graphs with $k=3$, $20$, $28$. We recorded them,
together with the scoring criteria, in version control before computing any exact ground state at those
densities (git commit \texttt{af2a3c8}, 2026-10-16 21:39 EDT; the prediction file has SHA-256
\texttt{b0818ffd\ldots e7dce81e}; see Data and Code Availability). We then computed exact ground states for
30 realisations per cell, at all five $\alpha$, along the 24-point $r_h$ grid ($25\,200$ optimisations).

\textbf{Certification.} All capped-BA and $k=3$ points and all random-regular points at $\alpha\le0.2$ are
certified. At $k=20$ and $28$ with $\alpha\ge0.5$ the solver often reaches the 120\,s limit: $91\%$--$98\%$
($k=20$) and $49\%$--$68\%$ ($k=28$) of the points within a factor $1.4$ of $r_h^*$ are certified;
uncertified points use the best incumbent. Near $r_h^*$, re-solving 28 uncertified points, sampled at random, with a 1200\,s limit returned the
same energy in 24 (10 now certified); two, both at $k=28$, $\alpha=0.5$, reached energies lower by at most
$0.5\%$ and raised the field share by $0.06$ and $0.16$, which would lower that cell's $r_h^*$ by about
$2\%$ and move its prediction error from $+2.7\%$ to about $+5\%$. At $r_h=0$, where $e_J(0)$ is measured, the
solver certifies every realisation of the main sweeps and of the new capped-BA, $k=3$ and $\alpha\le0.2$
cells, but only 2 of the 180 realisations at $k=20$ and $28$ with $\alpha\ge0.5$ (and 12 of the 1,080
realisations of the control families, Sec.~\ref{sec:families}). On all 210 uncertified realisations of the paper three
searches agree: the annealing used as the solver's start (200 reads of 4000 sweeps), the solver's incumbent after
120\,s, and heavy annealing (1000 reads of $20\,000$ sweeps, an algorithm unrelated to the solver) reach the
same energy. The solver's dual bounds there stay $3$--$45\%$ below, so these states are supported by the
agreement of two unrelated algorithms rather than proven optimal.

\begin{figure*}[t]
\centering
\includegraphics[width=\textwidth]{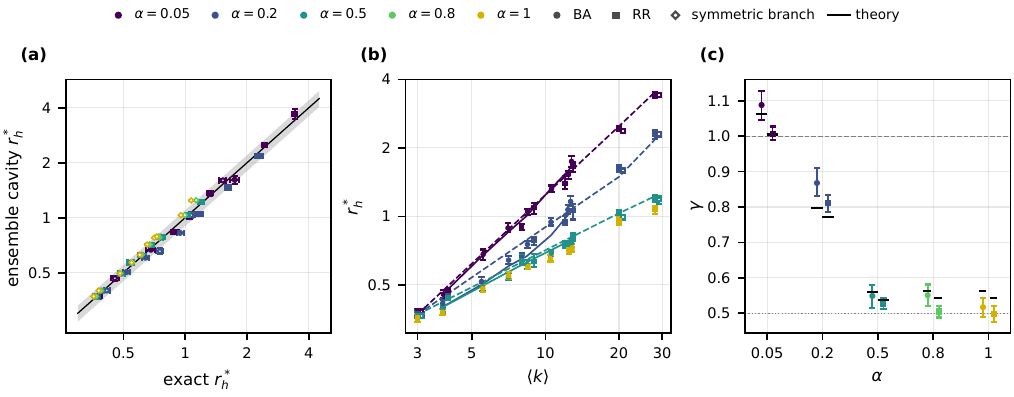}
\caption{Parameter-free ensemble cavity theory against exact ground states (30 realisations per cell).
In every panel circles are capped BA and squares random-regular graphs; colour is $\alpha$. (a) Predicted
against exact $r_h^*$ at $N=90$: filled, the 21 recorded predictions at seven new densities; open, the
densities of Table~\ref{tab:rstar}; diamonds, the sign-symmetric branch at $\alpha=0.8$ and $1$, computed
after the exact results. Error bars: exact, bootstrap 95\% CI; theory, standard deviation over four
seeds. Shaded, $\pm10\%$. (b) Exact $r_h^*$ against $\langle k\rangle$ (95\% CIs) at $N=90$ (filled) and,
for random-regular graphs, $N=180$ (open, shifted right by $4\%$ for visibility), with the theory evaluated at
the densities shown (lines; solid, capped BA; dashed, random-regular). (c) Exponent $\gamma$ of
$r_h^*\propto\langle k\rangle^\gamma$ (Table~\ref{tab:gamma}) with bootstrap 95\% CIs; solid bars, the
theory; dashed bars, its sign-symmetric branch (post hoc).}
\label{fig:ensemble}
\end{figure*}

\textbf{Crossover.} Table~\ref{tab:ensemble} and Fig.~\ref{fig:ensemble}(a) give the result. The predicted
$r_h^*$ lies within $10\%$ of the exact value in 19 of the 21 new cells, with median absolute error
$3.3\%$. The theory has its own Monte Carlo error. Over four population-dynamics seeds the standard deviation of
$r_h^*$ is below $2\%$ in 15 of the 21 cells but reaches $7\%$ at weak frustration and high density, where
the ordered branch ends in a jump whose position varies between runs: at $k=28$, $\alpha=0.05$ the four
seeds split between $3.30$ and $3.72$ (the recorded one is $3.72$), and a fourfold population gives
$3.30$. Elsewhere a fourfold population changes $r_h^*$ by at most $3\%$. Averaged over the seeds, the predictions are within $10\%$ in 20 of the 21
cells; the scoring above uses the recorded ones. We also changed the protocol of the population dynamics
in three ways, one at a time: four times as many sweeps; $r_h$ descending from $8$, so that the
field-dominated solution is followed down; and a fully ordered initial population. In 17 of the 21 cells
none of these moves $r_h^*$ by more than $5\%$. The exceptions are the three densest cells at $\alpha=0.05$
($m=15$, $k=20$, $k=28$), where the ordered solution ends in a jump and the protocols differ by up to
$12\%$ (at $k=20$ the descending branch gives a lower $r_h^*$ than the ascending one, so two solutions of the
cavity equations coexist there), and capped BA $m=7$ at $\alpha=0.2$, where longer runs raise $r_h^*$ by
$5$--$6\%$. Scored against the exact values, the three protocols place $r_h^*$ within $10\%$ in 21, 21 and 19
of the 21 cells, and at $\alpha=0.2$ all three keep every error negative (mean $-5.7\%$, $-5.2\%$ and
$-7.3\%$). In weakly frustrated dense networks the prediction is therefore no sharper than about $10\%$, the
size of the tolerance itself. The exact $r_h^*$ is resolved to $\pm3$--$5\%$, so the
10\% criterion fixed in advance is not a statistical one: the recorded prediction lies inside the exact
95\% interval in only 9 of the 21 cells, and within two combined standard errors in 14. The theory is accurate to a few
percent but not exact. Its errors are unsigned at $\alpha=0.05$ (mean $-0.5\%$) and $\alpha=0.5$ ($+2.1\%$),
but at $\alpha=0.2$ all seven are negative (mean $-6.7\%$), and both misses ($-12.5\%$ and $-10.0\%$, capped
BA $m=7$ and $15$) are there. The theory has $e_J(0)$ right at $\alpha=0.2$ (within $2.4\%$, unsigned), so the bias is
in where the field takes over, not in the coupling energy it must overcome: the predicted $r_h^*/e_J(0)$ is
$0.78$--$0.90$ against $0.85$--$0.98$ exactly. In the ensemble the ordered state yields to the field too
early. A post hoc diagnostic, made after the exact results, shows where. In units of $x=r_h/e_J(0)$ the
theory's crossover is early by $2$--$11\%$ in all seven $\alpha=0.2$ cells, and its field share rises more
abruptly than the exact one (width $w$ smaller by up to a factor three at high density, equal at $k=3$),
as it also does at $\alpha=0.05$, where the errors are unsigned; at $\alpha=1/2$ the widths agree to within $9\%$. We also computed the linear stability of the
replica-symmetric solution, $\lambda=\langle d\rangle_{\rm exc}P(|H|<|J|)$, the mean number of cavity biases
through which a small change of an incoming field propagates ($\lambda>1$: the solution is unstable). At
$\alpha=0.05$, $\lambda\le1.06$ at $r_h^*$ in all seven cells; at $\alpha=1/2$ the population is unordered
(population magnetisation $|m|\le0.02$) and $\lambda=1.7$--$2.7$ (0.65 at $k=3$), but there is no order for the field to
remove and the energy balance carries only the small replica-symmetric energy error. At $\alpha=0.2$ the
population is still partly ordered at $r_h^*$ ($|m|$ up to $0.5$) and $\lambda=1.5$--$1.7$ in the six cells
other than $k=3$, against $0.66$ at $k=3$, the cell with the smallest error. The bias thus appears where the
crossover is the loss of partial order in a regime where the replica-symmetric solution is itself unstable.
This is consistent with replica-symmetry breaking at intermediate frustration, though the diagnostic does
not prove it. This is not an effect of size or of short loops. We repeated the random-regular cells at $N=180$,
where the clustering of the $k=20$ and $28$ graphs is halved (30 realisations per cell; 10 at $k\ge20$,
$\alpha=0.5$, where $32$--$40\%$ of points are certified; elsewhere 12 of $5040$ points are uncertified): the exact
$r_h^*$ changes by at most $4\%$ in all nine cells, and at $\alpha=0.2$, $k=20$ and $28$ by $-2.0\%$ and
$0.0\%$, so the theory remains $8\%$ and $5\%$ low. The bias belongs to the replica-symmetric ensemble
solution itself.

\textbf{Coupling energy.} The recorded predictions of $e_J(0)$, taken at $r_h\to0$, and the exact values,
taken at $r_h=0$, agree to within $2.7\%$ in all 14 cells with $\alpha\le0.2$ (unsigned; Table~\ref{tab:ensemble}).
At $\alpha=0.5$ the theory's value is too large in every cell, including the fully certified ones, by $2.0\%$ at
$k=3$, $3.7$--$7.7\%$ on capped BA, and $5.4\%$ and $6.5\%$ at $k=20$ and $28$. It is not a finite-size effect:
at $N=180$ the exact $e_J(0)$ at $k=20$ and $28$ differs from that at $N=90$ by $+0.8\%$ and $+1.9\%$. It is the
known error of replica symmetry, which is below $1\%$ for the ground-state energy of a $\pm J$ spin glass at low
connectivity \cite{mezard2003,boettcher2003} and grows with connectivity towards its fully connected value,
$4.6\%$ (the replica-symmetric $\sqrt{2/\pi}=0.798$ against Parisi's $0.763$ \cite{parisi1980}).
The central relation of Sec.~\ref{sec:crossover} also holds at the new densities: the cell medians of the
per-realisation ratio $r_h^*/e_J(0)$ lie between $0.85$ and $1.03$ in all 35 new cells (all five $\alpha$),
against $0.85$--$1.00$ in the 36 cells of the main sweeps.

\textbf{Exponent.} Table~\ref{tab:gamma} gives $\gamma$ over the seven capped-BA densities
($\langle k\rangle=3.8$--$12.8$) and over random-regular graphs from $k=3$ to $28$, a ninefold range of
density [Fig.~\ref{fig:ensemble}(b,c)]. It is $1.01$--$1.09$ at $\alpha=0.05$ and $0.50$--$0.55$ at
$\alpha\ge0.5$, and at $\alpha=0.5$ it is unchanged ($0.52$ $[0.49,0.55]$) when the partly uncertified $k=20$
and $28$ cells are dropped. The predicted exponents lie inside the exact 95\% intervals at $\alpha=0.05$ and
$0.5$ and below them at $\alpha=0.2$, by $0.03$--$0.06$ with the same (mean-curve) definition for both.

\textbf{Curvature at intermediate frustration.} At $\alpha=0.2$ a single exponent is only a summary:
$r_h^*$ is not a power law of $\langle k\rangle$. On random-regular graphs its local slope between $k=3$ and
$20$ and between $k=20$ and $28$ is $0.77$ and $1.03$, and $\gamma=0.66$ $[0.62,0.69]$ over $k=3$--$12$
against $0.81$ over $k=3$--$28$. At $\alpha=0.05$ and $0.5$ the same slopes are $0.97$, $1.04$ and $0.55$,
$0.43$: no rise, so the curvature is specific to intermediate frustration. This is the deviation from a pure
mean-degree scaling that frustration produces, and it is a property of the zero-field ground state. Its
coupling energy relative to that of the fully aligned ferromagnet, $(1-2\alpha)(\langle k\rangle/2)
\langle|J|\rangle$, is $1.47$, $1.31$, $1.07$, $1.01$, $1.02$ and $0.99$ at $k=3$, $4$, $8$, $12$, $20$,
$28$, and its fraction of satisfied ties falls from $0.89$ to $1-\alpha=0.80$. In a sparse network the
ground state can satisfy some antagonistic ties at the price of a few cooperative ones; by
$\langle k\rangle\approx10$ this no longer pays: the ground state's coupling energy and satisfied-tie fraction
reach those of the aligned state, and its coupling energy grows like $\langle k\rangle$. This is the
ferromagnetic ordering of $\pm J$ models with ferromagnetically biased couplings becoming more complete as
connectivity grows \cite{viana1985,castellani2005}, read as a statement about ties. These are aggregate
measures; they do not show that every realisation is aligned configuration by configuration. The rising
slope is the approach to that energy. The ensemble theory
reproduces it without adjustment (relative coupling energy $1.48$ at $k=3$, $1.00$ at $k=20$; local slopes of
$r_h^*$ $0.73$ and $1.11$, averaged over seeds), and on capped BA graphs the same ratio falls from $1.30$ at $m=2$ to $1.01$
at $m=10$. The curvature is not a property of the small graphs: at $N=180$ the exact local slopes are
$0.77$ and $1.09$ at $\alpha=0.2$, against $0.96$, $1.08$ at $\alpha=0.05$ and $0.53$, $0.53$ at
$\alpha=0.5$ [open squares in Fig.~\ref{fig:ensemble}(b)].

\textbf{Strong frustration, $\alpha>1/2$.} The recorded predictions are for $\alpha\le1/2$, and the reason
we gave for this restriction was incomplete. Because $u(-J,H)=-u(J,H)$, the population update at $\alpha$
maps onto that at $1-\alpha$ with the cavity fields changing sign every generation, so the ferromagnetic
fixed point for $\alpha<1/2$ becomes a period-two orbit for $\alpha>1/2$. That is an antiferromagnetic
solution, which needs a bipartite graph, and unconstrained population dynamics converges to it (placing
$r_h^*$ too high by a factor $2.6$ at $m=10$, $\alpha=1$). The same symmetry has a further consequence:
on a sign-symmetric population, $P(H)=P(-H)$, the bias $u(-|J|,H)=u(|J|,-H)$ has the distribution of
$u(|J|,H)$, and the site and link terms are likewise invariant, so the update does not depend on $\alpha$ at
all. On a graph whose odd cycles exclude the antiferromagnetic branch, the relevant replica-symmetric
solution is this symmetric one, which is the solution at $\alpha=1/2$, and the theory predicts
$r_h^*(\alpha)=r_h^*(1/2)$ for every $\alpha>1/2$. We computed it after the exact results, so it is not part
of the recorded test, by symmetrising the cavity population every generation (the field distribution is
symmetric to within $4\%$ of $\langle|h|\rangle$). It changes $r_h^*$ at $\alpha=1/2$ by at most $0.5\%$
($1.7\%$ at $k=28$), gives the same $r_h^*$ at $\alpha=1$ as at $1/2$ (within $0.8\%$), and against the exact $r_h^*$ at $\alpha=0.8$ and $1$
on all ten densities it is within $10\%$ in 17 of 20 cells (median error $3.6\%$ at $\alpha=0.8$
and $6.7\%$ at $\alpha=1$, always too high) [diamonds in Fig.~\ref{fig:ensemble}(a)],
with exponents $0.56$ (BA) and $0.54$ (random-regular) against the exact $0.50$--$0.55$
(Table~\ref{tab:gamma}). The exact $r_h^*$ at $\alpha>1/2$ is $0.95$--$1.02$ ($\alpha=0.8$) and
$0.90$--$1.01$ ($\alpha=1$) times that at $\alpha=1/2$, lowest at the highest densities, a correction of up
to $10\%$ beyond the replica-symmetric ensemble. That an antiferromagnet on a sparse random graph has the
ground-state energy of a $\pm J$ spin glass, its long odd cycles frustrating it, is known
\cite{zdeborova2010,dembo2017}; it also explains why an expectation we recorded with the predictions, that
on sparse $k=3$ graphs $\alpha=1$ would be ferromagnet-like, failed [$r_h^*(1)/r_h^*(0.5)=0.97$].

\subsection{Does the outcome follow the aggregate field?}
\label{sec:outcome}

\begin{figure}[tbp]
\centering
\includegraphics[width=\columnwidth]{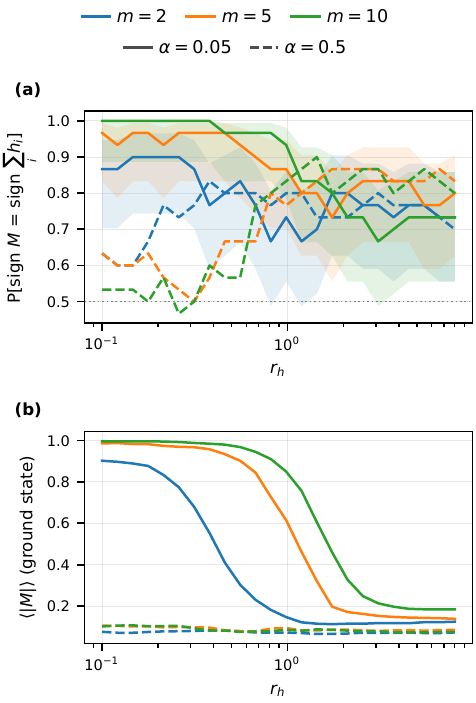}
\caption{(a) Fraction of realisations in which $\mathrm{sign}\,M=\mathrm{sign}\sum_ih_i$ in the exact
ground state, and (b) mean $|M|$, against $r_h$; $N=90$, 30 realisations per point, $\alpha=0.05$ (solid)
and $0.5$ (dashed). Dotted, chance.}
\label{fig:outcome}
\end{figure}

A socially interpretable consequence of the crossover is whether the collective outcome, the sign of the
ground-state vote margin $M$, follows the sign of the aggregate conviction $\sum_ih_i$
[Fig.~\ref{fig:outcome}(a)]. In weakly frustrated, dense networks the coupling amplifies the small
aggregate bias into a collective outcome: at $\alpha=0.05$ and $r_h\le0.2$ the outcome follows the
aggregate field in $30/30$ realisations at $m=10$ and in $28$--$29/30$ at $m=5$, and $|M|$ is close to one
[Fig.~\ref{fig:outcome}(b)]. Part of this is automatic: a near-consensus state lowers its field energy by
taking the sign of $\sum_ih_i$, so once the ground state is ordered its direction largely follows. What the
coupling adds is the ordering itself, a near-unanimous outcome; this is a ground-state idealisation of
strong social conformity, not a realistic vote margin. In frustrated networks ($\alpha\ge0.5$) the ground state is not ordered at
small $r_h$ (mean $|M|\le0.11$ for $r_h\le0.2$), and agreement with the aggregate field, $37$--$67\%$ of
realisations, is consistent with chance at this sample size.
As $r_h$ grows, every density approaches $73$--$90\%$, comparable to the $73$--$93\%$ of realisations in
which the headcount majority and the intensity-weighted aggregate $\sum_ih_i$ have the same sign. The
coupling-dominated, weakly frustrated regime is thus the one in which the network turns a narrow
intensity advantage into a near-certain collective outcome; with 30 realisations per point these
fractions carry binomial uncertainties of about $\pm0.15$, so we do not interpret finer structure.

\subsection{Exact low-energy landscape}
\label{sec:landscape}

\begin{figure}[t]
\centering
\includegraphics[width=\columnwidth]{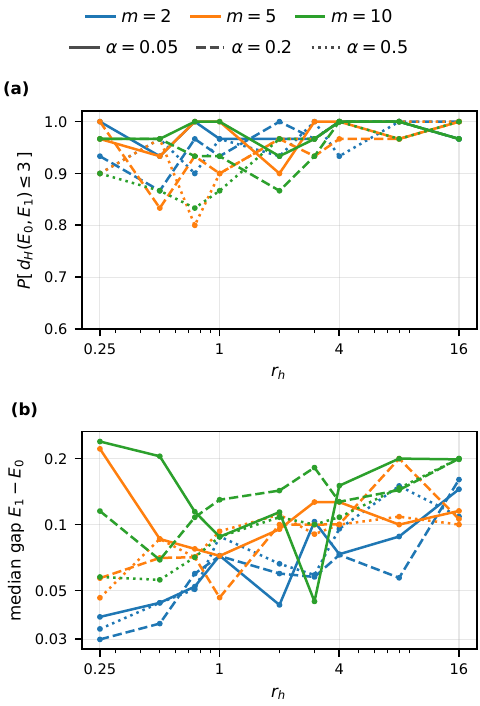}
\caption{Exact low-energy landscape ($N=90$, 30 realisations per cell; Gurobi solution pool, 20 lowest
configurations). (a) Fraction of instances whose lowest excited level lies within three spin flips of the
ground state; (b) median energy gap $E_1-E_0$. Colour, $m$; line style, $\alpha$.}
\label{fig:landscape}
\end{figure}

To characterise the low-energy landscape without dependence on a heuristic sampler, we enumerated the
20 lowest configurations of each instance with the Gurobi solution pool (PoolSearchMode~2) on 81 cells
($m\in\{2,5,10\}$, nine values of $r_h$ from $0.25$ to $16$, $\alpha\in\{0.05,0.2,0.5\}$; 30 realisations each,
$N=90$). The pool was proven complete within 60\,s in all but 2 of 810 instances at $m=2$, 68\% at $m=5$
and 19\% at $m=10$; otherwise it is the best set found and the first excited level $E_1$ is an upper
bound. Two guarantees should be kept apart: where the pool is proven complete, the 20 levels are certified
to be the lowest; elsewhere they are the lowest found. In every $(m,r_h,\alpha)$ cell the enumerated
configurations form a compact neighbourhood of the ground state, with no widely separated competing minima
among the resolved low-lying states. This does not exclude distant valleys at higher energy. Real signed social networks show both cases: one
has a single funnelled valley, another several near-isoenergetic valleys separated by barriers
\cite{facchetti2012}. It is what one
expects if, as in the
zero-temperature spin glass on random graphs, most spins carry the same local field in every
low-lying state \cite{perrupato2022}. The lowest excited level differs from
the ground state by one spin in the median instance of every cell, and by at most three spins in
$80$--$100\%$ of instances per cell [Fig.~\ref{fig:landscape}(a)]. Distant low-lying states are rare: in 5 of
2\,430 instances $E_1$ lies more than 30 spins from the ground state; four of these are the opposite
magnetisation branch at $\alpha=0.05$, $r_h\le0.5$. Ground-state degeneracy comes only from spins with
zero local field (32\%, 10\% and 6\% of instances at $m=2$, 5 and 10), and degenerate ground states
differ by at most five spins. The median gap $E_1-E_0$ is $0.03$--$0.24$ in energy units
[Fig.~\ref{fig:landscape}(b)], of the order of the smallest single-spin local field. Hamming distances between
low-lying samples of a stochastic annealer, which can end on either magnetisation branch at small $r_h$,
can therefore overstate the complexity of the lowest levels: the enumerated spectrum shows no competing
distant minima among them at these sizes. With gaps of the order of the smallest local field, 20 levels span
a narrow energy window, and the overlap distribution (Sec.~\ref{sec:pq}) shows that at finite temperature
frustrated networks do reach distant configurations.

\subsection{Overlap distribution}
\label{sec:pq}

We measured the overlap distribution at $\beta=2$ on $m=10$ graphs for $N=45$, $90$ and $180$ (60, 40
and 24 realisations), at $r_h=0.25$ (field share $\approx0.1$--$0.3$, coupling-dominated) and $r_h=1$
(field-dominated for $\alpha\ge0.2$). For reference, the zero-field Bethe estimate of the spin-glass
temperature of this coupling distribution, $(\kappa-1)\langle\tanh^2\beta_cJ\rangle=1$, is $T_c\approx1.2$,
so $\beta=2$ is well below it. All 30 cells pass the equilibration test: the disorder-averaged energy in
the last two logarithmic bins agrees to better than $10^{-4}$ relative, and $\langle q^2\rangle$ to within
$1.2\%$. Every realisation made at least 51 round trips of the temperature ladder, with swap acceptance at
least $0.14$.

\begin{figure*}[t]
\centering
\includegraphics[width=\textwidth]{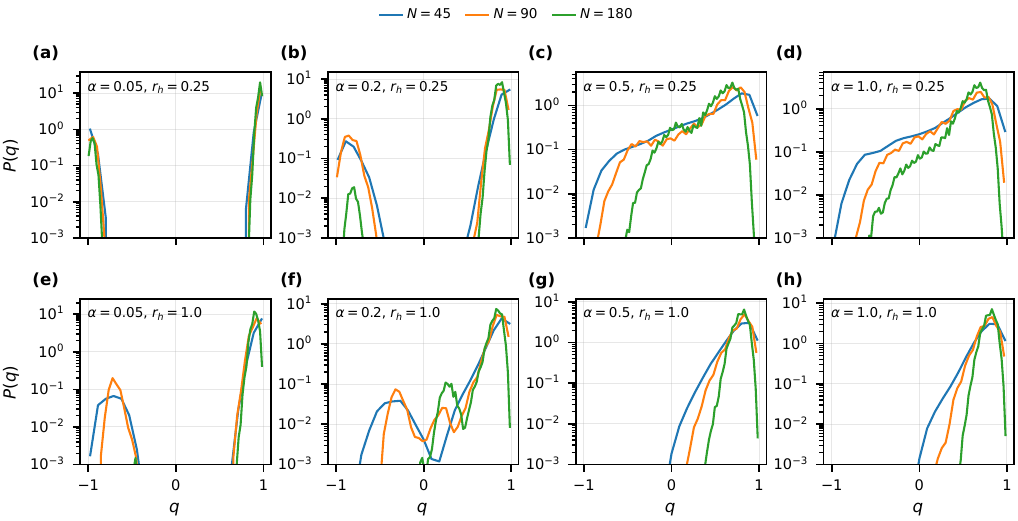}
\caption{Signed overlap distribution $P(q)$ at $\beta=2$ from parallel tempering, $m=10$, for $N=45$, $90$,
$180$ (pooled over realisations; bins of width $4/N$, matching the $2/N$ lattice of overlaps); top row
$r_h=0.25$, bottom row $r_h=1$.}
\label{fig:pq}
\end{figure*}

\begin{figure}[t]
\centering
\includegraphics[width=\columnwidth]{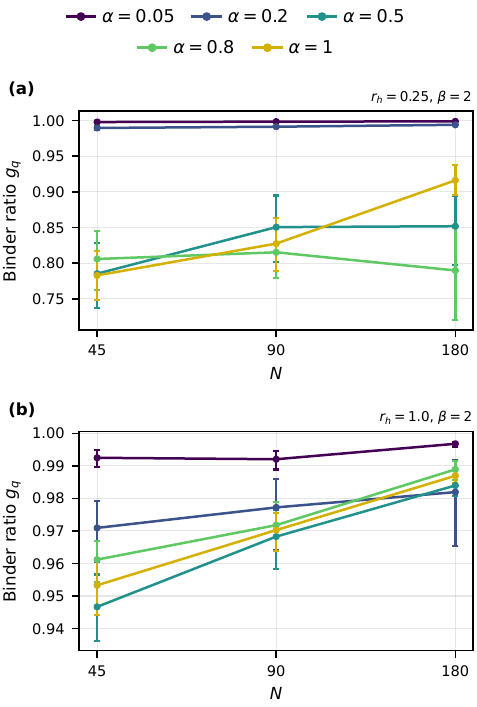}
\caption{Binder ratio $g_q=\frac12(3-\langle q^4\rangle/\langle q^2\rangle^2)$ of the overlap at $\beta=2$
against $N$, by $\alpha$, for (a) $r_h=0.25$ and (b) $r_h=1$; bootstrap 95\% CIs over realisations.}
\label{fig:binder}
\end{figure}

The overlap distributions fall into two classes (Fig.~\ref{fig:pq}). In weakly frustrated networks
($\alpha\le0.2$) $P(q)$ has two sharp peaks near $q=\pm q_{\rm EA}$ with $q_{\rm EA}\approx0.9$--$0.95$: the
two chains sit in the same or in opposite magnetisation branches. The weight at $q<0$ falls with $N$ and
with $r_h$ as the aggregate field selects one branch, and $g_q\ge0.97$ throughout. This is a ferromagnet in
a weak random field, not a spin glass. In frustrated networks ($\alpha\ge0.5$) $P(q)$ has a single peak at
$q\approx0.7$--$0.8>0$ with a tail towards small and negative $q$. At $r_h=1$ the tail vanishes with $N$
and $g_q$ rises towards 1 [Fig.~\ref{fig:binder}(b)], as expected for a replica-symmetric state polarised by
the field. At $r_h=0.25$ the tail is substantial: the weight at $q<0.2$ is $0.19\to0.13\to0.11$
($\alpha=0.5$), $0.15\to0.14\to0.16$ ($\alpha=0.8$) and $0.18\to0.12\to0.04$ ($\alpha=1$) from $N=45$ to
$180$. The Binder ratio \cite{binder1981}, $0.78$--$0.92$, shows no consistent trend with $N$ [Fig.~\ref{fig:binder}(a)].
These sizes and a single temperature cannot decide whether frustrated networks at small $r_h$ have a
broken-replica-symmetry phase in a field, a question that remains open even for spin glasses on random
graphs in a field \cite{young2004,katzgraber2001}, and we make no such claim: the overlap data are
descriptive and do not identify a phase. What the data do show complements Sec.~\ref{sec:landscape}: the
lowest levels are clustered around the ground state, but where the ground state is coupling-dominated and frustrated,
thermal fluctuations at $T\approx0.4\,T_c$ still explore configurations far from the ground state. In
weakly frustrated networks they explore only the two magnetisation branches.

\section{Discussion}

The strongest evidence in this paper is a prediction: a zero-temperature cavity theory with no fitted
parameter, recorded before the exact ground states were computed, placed the crossover within $10\%$ in
19 of 21 cells at seven densities it had not seen, with the known replica-symmetric error in $e_J(0)$ at
$\alpha=1/2$ and a systematic underestimate of $r_h^*$ at $\alpha=0.2$. That underestimate is the most
informative failure in the paper. At $\alpha=0.2$ the theory gets the coupling energy right but lets the field
take over too early and too abruptly, and it does so where its replica-symmetric solution is linearly unstable
while the population is still partly ordered (at $\alpha=1/2$ the solution is also unstable, but there is no
order for the field to remove, and the error stays small); this points to replica-symmetry breaking in a random field at
intermediate frustration, an open question that our diagnostic suggests but does not settle. The physics behind
the crossover itself is simple. The field--coupling crossover of this model sits where the field energy matches the coupling energy of
the zero-field ground state. For a model of an electorate this makes the question ``how strong must
individual conviction be, relative to social pressure, before it decides the outcome?'' a question about
the ground-state energy of the social network's coupling term alone, a quantity fixed by the signed network
before any opinion is known. Within the model, and with the caveats below (stylised couplings, ground states
rather than dynamics), this has four consequences. First, when almost all ties are
cooperative, that energy grows with the number of ties, and denser networks need proportionally stronger
convictions to override them. When a substantial fraction of ties is antagonistic, the network cannot
satisfy its own ties, the coupling energy grows only about as the square root of the number of ties over
the densities studied, and density matters much less: going from $\langle k\rangle=3.8$ to $12.4$ raises $r_h^*$ by $3.4\times$ at
$\alpha=0.05$ but by only $1.9\times$ at $\alpha\ge0.5$. Adding ties to a polarised network adds little
social pressure, because the new ties partly cancel. Because the scaling is the same on
Erd\H{o}s--R\'enyi, random-regular and configuration-model graphs, it is a property of mean degree and
frustration, not of the scale-free degree distribution; heterogeneity only broadens the crossover, because
the best-connected voters, held by the most ties, are the last to vote their own conviction
(Sec.~\ref{sec:families}). Second,
at intermediate frustration the answer depends on density in a further way (Sec.~\ref{sec:ensemble}): a
sparse network accommodates its antagonistic ties, its ground state satisfying some of them at the price of
a few cooperative ones, so that its coupling energy exceeds that of the aligned state by up to $47\%$ at
$k=3$; by $\langle k\rangle\approx10$ this no longer pays, and on aggregate antagonism acts only as a uniform
discount $(1-2\alpha)$ on cooperation. Third, cooperative networks turn a small aggregate bias into a
near-consensus outcome at small $r_h$, and frustration removes this (Sec.~\ref{sec:outcome}). Fourth, the
graph itself smooths the crossover: population mean field, in which every voter feels the same social pull,
predicts an abrupt collapse of consensus, the defection of the most committed minority voters triggering
everyone else's, whereas on the actual graphs, where the pull differs from voter to voter and the fields are
broad compared with the couplings, the field share
rises gradually at every size studied (Sec.~\ref{sec:theory}). The crossover is
not a polarisation transition of the kind found in Ref.~\cite{korbel2026}, which has hysteresis and sets in
above a critical campaign spending: it is a smooth energy balance that does not sharpen over the sizes
studied, located by the zero-field coupling energy.

Methodologically, three points generalise. First, certified ground states are cheap at the sizes used in
much computational sociophysics, and they settle questions that annealing output cannot: here they
confirmed every native-schedule SA ground state of the earlier regime map and exposed the failures of
an energy-matched schedule at large $r_h$. Second, population-level mean field, the usual theoretical
companion of such models \cite{brock2001,michard2005}, has no way to represent the coupling energy of a
frustrated network. The zero-temperature cavity method, by contrast, predicts the crossover to within about
$10\%$, and within a few percent in most cells. On the actual graph it does so even where its decoding is imperfect. On the graph ensemble it
needs no instance and no fitted parameter, and its predictions held to within $10\%$ at densities they had
not seen, with a small systematic bias at intermediate frustration; its sign-symmetric branch, the
consequence of the gauge symmetry of the cavity biases, also gives the crossover of strongly frustrated
networks. Third, population statistics of
annealer output depend on the schedule (Appendix~\ref{app:schedule}), and distances between low-lying
annealer samples can suggest a rugged landscape where exact enumeration finds the lowest levels
clustered around the ground state (Sec.~\ref{sec:landscape}); ground-state statistics, once certified, do not have these problems.

\textbf{Limitations.} Coupling magnitudes are stylised, so the location of a real electorate on the $r_h$
axis is uncertain by at least the uncertainty in the units of $J$; the results are stated as functions of
$r_h$ for that reason. Bond signs are i.i.d.; structured sign placement is treated in the companion paper
\cite{romero2026row1}. Exact ground states are limited to $N\le180$ by the solver licence available, and
the larger sizes rely on a heuristic validated at $N=180$; the densest random-regular graphs ($k=20$, $28$)
are not locally tree-like at $N=90$ (clustering $\approx(k-1)/N\approx0.2$--$0.3$), and we checked them at
$N=180$ only. The overlap distribution is computed at one
density and two values of $r_h$ at a single temperature, and three sizes are too few for a finite-size
analysis of a possible spin-glass phase in a field \cite{dealmeida1978,young2004}, so we make no claim about
one. The model is solved for its global energy minimum. That every voter's choice is then a best response
to the others does not imply that an electorate reaches this configuration: real opinion dynamics can stall
in metastable states, and whether, and how fast, a population of voters finds the ground state is a dynamical
question this paper does not address. The out-of-sample test is a test within the model: the theory and
the exact ground states share the model's field and coupling distributions, so it shows that the cavity
theory captures the model, not that the model describes an electorate; and its $10\%$ criterion is a
practical tolerance fixed in advance, not a statistical test, comparable in the densest weakly frustrated
cells to the theory's own dependence on its numerical protocol. Real political-discussion networks have smaller cores than our densest graphs; the $\langle
k\rangle$ and $\langle k\rangle^{1/2}$ scalings, not the particular densities, are the transferable result.

\section*{Data and Code Availability}

The simulation, exact-solver and analysis code, and the scripts that produce every figure and table, are at
\url{https://github.com/RomeroNatalia/17506376}. The prediction file of Sec.~\ref{sec:ensemble} is in the same
repository (\texttt{p12/pred\_preregistered.json}, SHA-256 \texttt{b0818ffd\ldots e7dce81e}); it was recorded under
commit \texttt{af2a3c8} (2026-10-16) of the author's development repository, before the exact ground states at those
densities were computed, and that commit record is available on request, as are the per-realisation results with
their seeds. An archived version with an assigned DOI will accompany journal publication.

\appendix

\section{Annealing schedule conventions}
\label{app:schedule}

Scaling $h$ up by $r_h$ at fixed $J$ and scaling $J$ down by $r_h$ at fixed $h$ give Hamiltonians related
by $H_{h}=r_hH_{J}$: they share every configuration's ranking, hence the ground state, while energies
differ by the factor $r_h$. Ratio observables such as $f$ are identical. For annealing, however, the
schedule's $\beta$ range is in absolute units, so a fixed range applied to $H_h$ is equivalent to the range
divided by $r_h$ applied to $H_J$. The native convention keeps $\beta\in[0.1,10]$ at every $r_h$; the
adjusted one divides it by $r_h$. Ground states agree whenever the annealer finds them
(Sec.~\ref{sec:exact}). Statistics of the sampled population do not: the fraction of reads that land in the
lowest configuration found rises with $r_h$ under the native convention and falls to zero under the
adjusted one (Spearman $\rho=+0.18$ to $+0.62$ versus $-0.68$ to $-0.86$ across the 15 $(m,\alpha)$ series),
because at large $r_h$ the adjusted schedule ends at a temperature high compared with $|J|$, so the
zero-field voters remain thermally disordered. Such population statistics measure the schedule rather
than the landscape, and none is used in the main text.

\section{Percolation of the signed subgraphs}
\label{app:percolation}

\begin{figure*}[t]
\centering
\includegraphics[width=\textwidth]{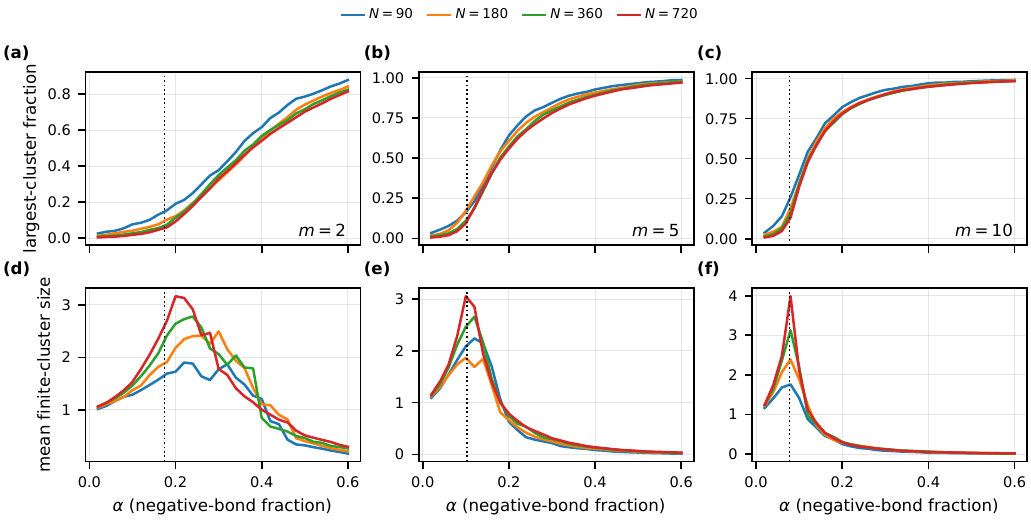}
\caption{Bond percolation of the negative-bond subgraph on capped BA graphs. (a--c) Largest-cluster
fraction and (d--f) mean finite-cluster size against $\alpha$, for $N=90$--$720$; dotted, $p_c=1/(\kappa-1)$.}
\label{fig:perc}
\end{figure*}

The negative-bond subgraph is a bond-percolation process on the graph at occupation $\alpha$, with
Molloy--Reed threshold $p_c=1/(\kappa-1)$ \cite{molloy1995}. Figure~\ref{fig:perc} shows its largest-cluster
fraction and mean finite-cluster size for $N=90$--$720$ (20 graphs per point). The finite-cluster size peaks at
$p_c$ and the peak grows with $N$ at every density, the standard finite-size signature of the percolation
transition \cite{stauffer1994}. Percolation of the antagonistic subgraph does not itself mark a change in
the ground-state observables above. The thresholds, $p_c=0.17$, $0.10$ and $0.07$ for $m=2$, $5$, $10$ at $N=90$
($0.17$, $0.10$ and $0.08$ at $N=720$), all lie between $\alpha=0.05$ and $0.2$, so our $\alpha$ grid cannot
resolve the crossover near $p_c$; but the density exponent $\gamma$ of Sec.~\ref{sec:crossover}, fitted across
the three densities, keeps falling well above $p_c$ ($1.05$, $0.80$ and $0.56$ at $\alpha=0.05$, $0.2$ and $0.5$).

\section{A rigorous window for the crossover}
\label{app:bounds}

Write the energy of a configuration $\bm s$ at field ratio $r$ as $E=E_h+E_J$, with
$E_h=-r\sum_ih_is_i$ and $E_J=-\sum_{(i,j)}J_{ij}s_is_j$, and let $e_J(0)=\max_{\bm s}(-E_J)/N$ be the
coupling energy per spin of the zero-field ground state. At the crossover $r=r_h^*$ the ground state $\bm
s^*$ has $|E_h|=|E_J|$; both terms are negative there in every realisation of this paper
(Sec.~\ref{sec:observables}), so $E_h(\bm s^*)=E_J(\bm s^*)=-Nc$ with $c>0$, and $E(\bm s^*)=-2Nc$.

\emph{Lower bound.} Let $\bm s_0$ be the zero-field ground state, globally flipped if necessary so that
$\sum_ih_is_{0,i}\ge0$. Then $E(\bm s_0)\le-Ne_J(0)$, and optimality of $\bm s^*$ gives $-2Nc\le-Ne_J(0)$,
i.e. $c\ge e_J(0)/2$. Since $Nc=r_h^*|\sum_ih_is^*_i|\le r_h^*N\langle|h|\rangle$ with
$\langle|h|\rangle=\sum_i|h_i|/N$, $r_h^*\ge e_J(0)/(2\langle|h|\rangle)$.

\emph{Upper bound.} Let $\bm s_h$ satisfy $s_{h,i}=\mathrm{sign}\,h_i$ wherever $h_i\neq0$ (zero-field spins
arbitrary), and $\varepsilon_h=E_J(\bm s_h)/N$. Then $E(\bm s_h)=-r_h^*N\langle|h|\rangle+N\varepsilon_h$, and
optimality gives $r_h^*\langle|h|\rangle\le2c+\varepsilon_h$. By definition of $e_J(0)$, $c\le e_J(0)$, so
$r_h^*\langle|h|\rangle\le2e_J(0)+\varepsilon_h$. Together these are Eq.~(\ref{eq:window}).

\emph{Check.} For every realisation we took $r_h^*$ from its own field-share curve, $e_J(0)$ from the
zero-field solve (Sec.~\ref{sec:observables}), and $\bm s_h$ with the zero-field spins set greedily to lower
$E_J$. All 3,440 lie inside the window, including the 210 whose $e_J(0)$ is not certified; $r_h^*\langle|h|\rangle/e_J(0)$
has median $1.00$ and interquartile range $0.94$--$1.05$, and $\varepsilon_h/e_J(0)$ ranges from $-0.53$ to
$0.44$ (median $-0.07$).

\section{Simulated-annealing settings against the exact optimum}
\label{app:sa}

We tested dwave-neal SA (geometric $\beta$ schedule from $\beta_{\min}=0.1$) against the certified optimum
on 60 independent $N=90$ instances [$m\in\{5,10\}$, $\alpha\in\{0.2,0.5\}$, $r_h\in\{0.25,1,4\}$, five
realisations each] for 18 combinations of final inverse temperature $\beta_{\max}$, sweeps per read and
reads (Table~\ref{tab:sa}). With $\beta_{\max}\ge30$ every setting, down to 100 reads of 1500 sweeps,
found the exact ground state in all 60 instances. At $\beta_{\max}=10$, 100 reads missed it in one
($1500$ sweeps) and two ($5000$ sweeps) instances at $r_h\le1$, with relative energy error at most
$1.5\times10^{-3}$; 1000 reads, the setting of the original regime map, always succeeded. The heuristic
used here for $N=360$ and $720$ (200 reads of 4000 sweeps, $\beta_{\max}=100$) lies inside the
always-successful region at $N=90$ and is validated against exact optima at $N=180$ in
Sec.~\ref{sec:exact}.

\begin{table}[h]
\caption{Number of the 60 test instances ($N=90$) in which SA found the certified ground state, by
final inverse temperature $\beta_{\max}$, sweeps per read and number of reads.}
\label{tab:sa}
\begin{ruledtabular}
\begin{tabular}{lcccccc}
 & \multicolumn{2}{c}{1500 sweeps} & \multicolumn{2}{c}{5000 sweeps} & \multicolumn{2}{c}{20000 sweeps}\\
$\beta_{\max}$ & 100 & 1000 & 100 & 1000 & 100 & 1000 \\
\hline
$10$ & 59 & 60 & 58 & 60 & 60 & 60 \\
$30$ & 60 & 60 & 60 & 60 & 60 & 60 \\
$100$ & 60 & 60 & 60 & 60 & 60 & 60 \\
\end{tabular}
\end{ruledtabular}
\end{table}

\section{The crossover and voter switching}
\label{app:switching}

\begin{figure}[htb]
\centering
\includegraphics[width=\columnwidth]{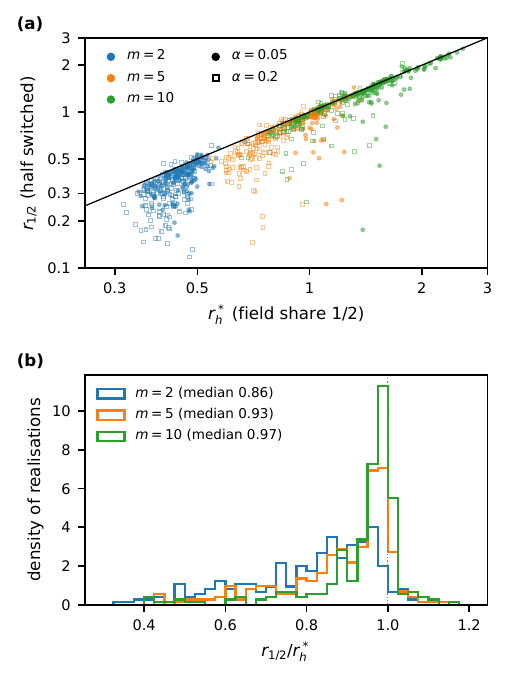}
\caption{The energy crossover and voter switching (exact ground states, $N=90$, five families, $\alpha\le0.2$,
30 realisations per cell). (a) $r_{1/2}$, at which half of the voters held against their own field have
switched, against $r_h^*$, one point per realisation; the line is $r_{1/2}=r_h^*$. (b) Distribution of
$r_{1/2}/r_h^*$ by density, pooled over families.}
\label{fig:switching}
\end{figure}

For the 900 realisations of Sec.~\ref{sec:families} with $\alpha\le0.2$ we took the partisans and leaners
($|h_i|\ge1$) whose ground-state opinion at the smallest $r_h$ is against their own field and recorded, for each,
the smallest grid value of $r_h$ from which it agrees with its field at every larger one. $r_{1/2}$ is the $r_h$,
interpolated in $\ln r_h$, at which half of them have switched; $r_h^*$ is taken from the same realisation's
field-share curve. Figure~\ref{fig:switching}(a) compares the two realisation by realisation. The ratio
$r_{1/2}/r_h^*$ has median $0.93$ (interquartile range $0.81$--$0.98$); pooled over families it has median
$0.86$, $0.93$ and $0.97$ at $m=2$, $5$ and $10$ [Fig.~\ref{fig:switching}(b)], and cell medians are
$0.75$--$0.91$ at $m=2$ and $0.92$--$0.99$ at $m=10$. The tail towards small ratios (down to $0.13$) comes mostly from the sparsest networks
[$m=2$, Fig.~\ref{fig:switching}(b)], in which half of the held-back voters switch well before the field
carries half of the energy. In dense networks the energy crossover and the point at which
the typical held-back partisan changes opinion coincide to within a few percent.

\section{Graph wiring and the hub effect}
\label{app:wiring}

\begin{figure}[htb]
\centering
\includegraphics[width=\columnwidth]{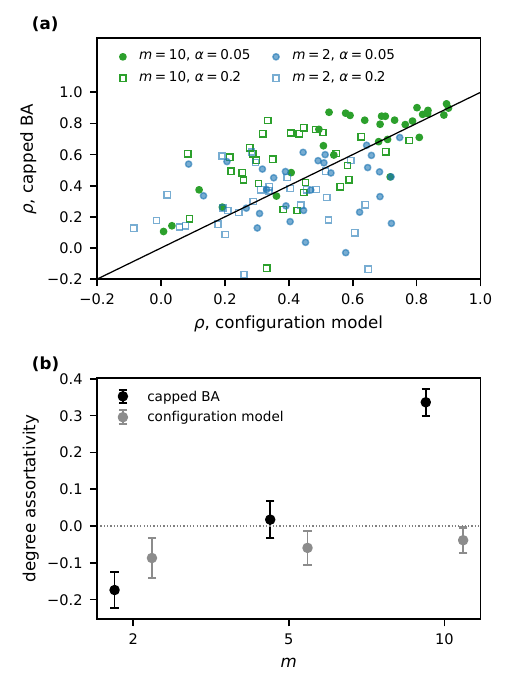}
\caption{Wiring and the hub effect ($N=90$, 30 realisations per cell). (a) Spearman $\rho$ between degree and
switching field energy on capped BA against the configuration model built on the same degrees, fields and
bond signs, one point per realisation, for $m=10$ and $m=2$; the line is equality. (b) Degree assortativity of
the two families by density (mean and standard deviation over realisations).}
\label{fig:wiring}
\end{figure}

The capped BA graphs and the configuration-model graphs of Sec.~\ref{sec:families} have the same degree sequence,
fields and bond signs, realisation by realisation, so they can be compared in pairs. The rank correlation $\rho$
between a voter's degree and the field energy at which it switches (Sec.~\ref{sec:families}) is the same on the
two families at $m=2$ and $5$ (paired Wilcoxon $p\ge0.13$), but larger on capped BA at $m=10$: median $0.81$
against $0.69$ at $\alpha=0.05$ (larger in 24 of 30 pairs, $p=1.5\times10^{-4}$) and $0.58$ against $0.39$ at
$\alpha=0.2$ (20 of 30, $p=3.5\times10^{-3}$) [Fig.~\ref{fig:wiring}(a)]. The difference is in the wiring. The
degree cap removes an over-cap node's ties to its least-connected neighbours, so at $m=10$, where it removes
about $30\%$ of the edges, the remaining hubs are tied mainly to one another: the degree assortativity is $0.34$
on capped BA against $-0.04$ on the configuration model, while at $m=2$ and $5$, where the cap removes few
edges, both are near zero or negative [Fig.~\ref{fig:wiring}(b)]. A voter's mean neighbour degree delays its
switch by a similar small amount on both families (partial Spearman coefficient, controlling for its own
degree, $0.11$--$0.16$ at $\alpha=0.05$ on both); on capped BA at $m=10$ this effect points the same way as the voter's own degree, which
strengthens the correlation with degree. This is an explanation consistent with these three observations, not a
controlled test, which would require rewiring the configuration-model graphs to a prescribed assortativity. It
is also a side effect of the cap worth knowing when capped BA graphs are used as models of social networks.

\end{document}